\documentclass[twocolumn]{aastex62} 
\usepackage{graphicx}
\usepackage{txfonts}
\usepackage{threeparttable}

\received{xxx}
\revised{xxx}
\accepted{}
\submitjournal{ApJ}
\shorttitle{Broadband X-ray variability in NGC 4395}
\shortauthors{Kammoun et al.}

\newcommand{\xmm}{\textit{XMM-Newton}}
\newcommand{\nustar}{\textit{NuSTAR}}

\begin{document} 

\title{The nature of the broadband X-ray variability in the dwarf Seyfert galaxy NGC 4395}
\correspondingauthor{E. Kammoun}
\email{ekammoun@umich.edu}

\author[0000-0002-0273-218X]{E. S. Kammoun}
\affiliation{Department of Astronomy, University of Michigan, 1085 South University Avenue, Ann Arbor, MI 48109-1107, USA}

\author[0000-0001-9226-8992]{E. Nardini}
\affiliation{INAF - Osservatorio Astrofisico di Arcetri, Largo Enrico Fermi 5, I-50125 Firenze, Italy }

\author[0000-0002-0572-9613]{A. Zoghbi}
\affiliation{Department of Astronomy, University of Michigan, 1085 South University Avenue, Ann Arbor, MI 48109-1107, USA}

\author{J. M. Miller}
\affiliation{Department of Astronomy, University of Michigan, 1085 South University Avenue, Ann Arbor, MI 48109-1107, USA}

\author[0000-0002-8294-9281]{E. M. Cackett}
\affiliation{Department of Physics \& Astronomy, Wayne State University, 666 W. Hancock Street, Detroit, MI 48201, USA}

\author{E. Gallo}
\affiliation{Department of Astronomy, University of Michigan, 1085 South University Avenue, Ann Arbor, MI 48109-1107, USA}

\author{M. T. Reynolds}
\affiliation{Department of Astronomy, University of Michigan, 1085 South University Avenue, Ann Arbor, MI 48109-1107, USA}

\author[0000-0002-3556-977X]{G. Risaliti}
\affiliation{Dipartimento di Fisica e Astronomia, Universit\`{a} di Firenze, Via G. Sansone 1, 50019 Sesto Fiorentino (Firenze), Italy}
\affiliation{INAF - Osservatorio Astrofisico di Arcetri, Largo Enrico Fermi 5, I-50125 Firenze, Italy }

\author{D. Barret}
\affiliation{IRAP, Universit\'{e} de Toulouse, CNRS, UPS, CNES, 9, Avenue du Colonel Roche, BP 44346, 31028 Toulouse Cedex 4, France}

\author{W. N. Brandt}
\affiliation{Department of Astronomy and Astrophysics, 525 Davey Lab, The Pennsylvania State University, University Park, PA 16802, USA}
\affiliation{Institute for Gravitation and the Cosmos, The Pennsylvania State University, University Park, PA 16802, USA}
\affiliation{Department of Physics, 104 Davey Lab, The Pennsylvania State University, University Park, PA 16802, USA}

\author{L. W. Brenneman}
\affiliation{Harvard-Smithsonian Center for Astrophysics, 60 Garden St., Cambridge, MA 02138, USA}

\author{J. S. Kaastra}
\affiliation{SRON Netherlands Institute for Space Research, Sorbonnelaan 2, 3584 CA Utrecht, the Netherlands}
\affiliation{Leiden Observatory, Leiden University, PO Box 9513, 2300 RA Leiden, the Netherlands}

\author{M. Koss}
\affiliation{Eureka Scientific, 2452 Delmer Street Suite 100, Oakland, CA 94602-3017, USA}

\author{A. M. Lohfink}
\affiliation{Montana State University, P.O. Box 173840, Bozeman, MT 59717-3840, USA}

\author{R. F. Mushotzky}
\affiliation{Department of Astronomy and Joint Space-Science Institute, University of Maryland, College Park, MD 20742, USA}

\author{J. Raymond}
\affiliation{Harvard-Smithsonian Center for Astrophysics, 60 Garden St., Cambridge, MA 02138, USA}

\author{D. Stern}
\affiliation{Jet Propulsion Laboratory, California Institute of Technology, 4800 Oak Grove Drive, MS 169-221, Pasadena, CA 91109, USA}

\begin{abstract}

We present a flux-resolved X-ray analysis of the dwarf Seyfert 1.8 galaxy NGC~4395, based on three archival \xmm\ and one archival \nustar\ observations. The source is known to harbor a low mass black hole ($\sim 10^4- {\rm a~ few~}\times 10^{5}~\rm M_\odot$) and shows strong variability in the full X-ray range during these observations. We model the flux-resolved spectra of the source assuming three absorbing layers: neutral, mildly ionized, and highly ionized ($N_{\rm H} \sim 1.6\times 10^{22}-3.4 \times 10^{23}~\rm cm^{-2}$, $\sim 0.8-7.8 \times 10^{22}~\rm cm^{-2}$, and $ 3.8 \times 10^{22}~\rm cm^{-2}$, respectively. The source also shows intrinsic variability by a factor of $\sim 3$, on short timescales, due to changes in the nuclear flux, assumed to be a power law ($\Gamma = 1.6-1.67$). Our results show a positive correlation between the intrinsic flux and the absorbers' ionization parameter. The covering fraction of the neutral absorber varies during the first \xmm\ observation, which could explain the pronounced soft X-ray variability. However, the source remains fully covered by this layer during the other two observations, largely suppressing the soft X-ray variability. This suggests an inhomogeneous and layered structure in the broad line region. We also find a difference in the characteristic timescale of the power spectra between different energy ranges and observations. We finally show simulated spectra with {\it XRISM}, {\it eXTP}, and {\it Athena}, which will allow us to characterize the different absorbers, study their dynamics, and will help us identify their locations and sizes.

\end{abstract}


%
\section{Introduction}
\label{sec:intro}

The dwarf Seyfert 1.8 galaxy NGC 4395 \citep[$d = 4.2~\rm Mpc$][]{Karachentsev98} is one of the most X-ray variable non-jetted active galactic nuclei \citep[AGN; e.g.,][]{Iwasawa00, Vaughan05, Iwasawa10}. The optical and ultraviolet (UV) spectra of this source show high-ionization forbidden lines with broad wings corresponding to gas velocities larger than $\sim 10^3~\rm km~s^{-1}$ \citep{Filippenko89}, in addition to permitted lines such as \ion{C}{4}, \ion{Mg}{2}, \ion{O}{3} and H$\alpha$ \citep[see e.g.,][]{Filippenko93}. \cite{Peterson05} obtained a mass of $M_{\rm BH} = (3.6 \pm 1.1)\times 10^5~\rm M_\odot$, based on reverberation mapping of \ion{C}{4}. More recently, \cite{Woo19} estimated the time delay in the H$\alpha$ band to be $83\pm 14$ minutes and found a small velocity dispersion of $\sigma_{\rm H\alpha} = 426 \pm 1 ~\rm km~s^{-1}$, inferring an even lower mass of $M_{\rm BH} = 9.1_{-1.6}^{+1.5} \times 10^3~\rm M_\odot$. The source can thus be considered to lie at the highest end of the still elusive intermediate mass black hole population \citep[see e.g.,][and references therein]{Koliopanos17,Mezcua18}, representing a scaled-down (by $\sim 2$ orders of magnitude) version of ordinary and more luminous Seyfert galaxies.

\begin{figure*}
\centering

\includegraphics[width = 1.\linewidth]{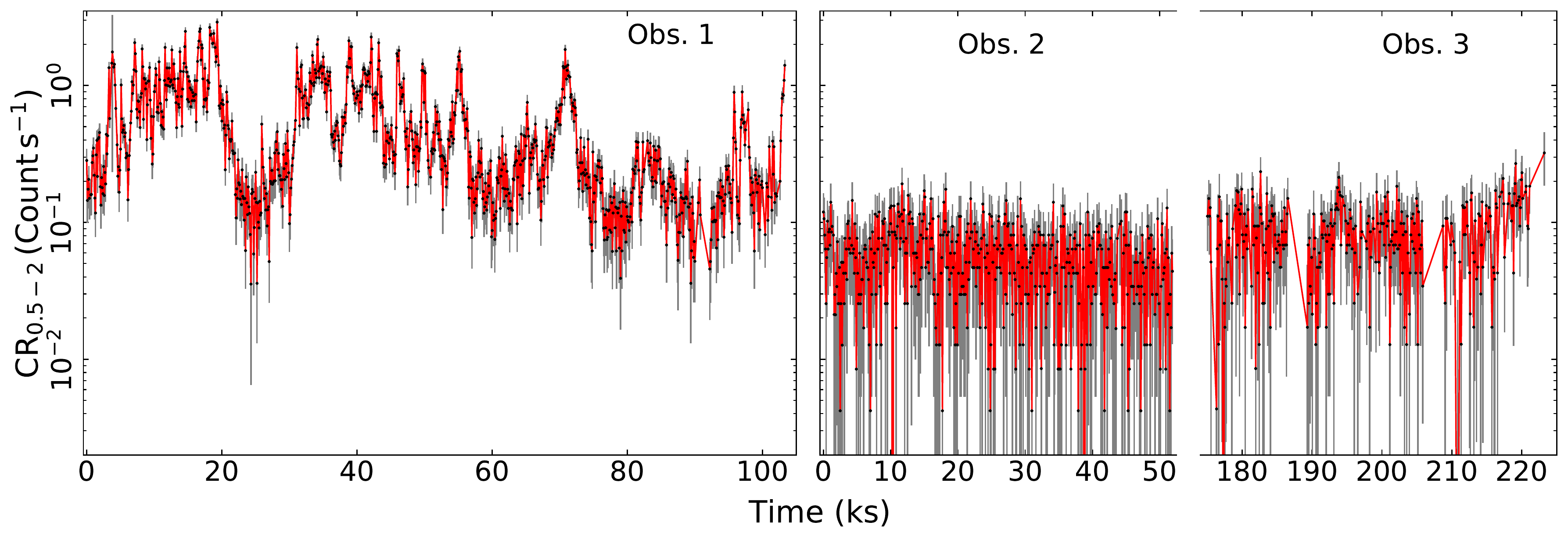}
\includegraphics[width = 1.\linewidth]{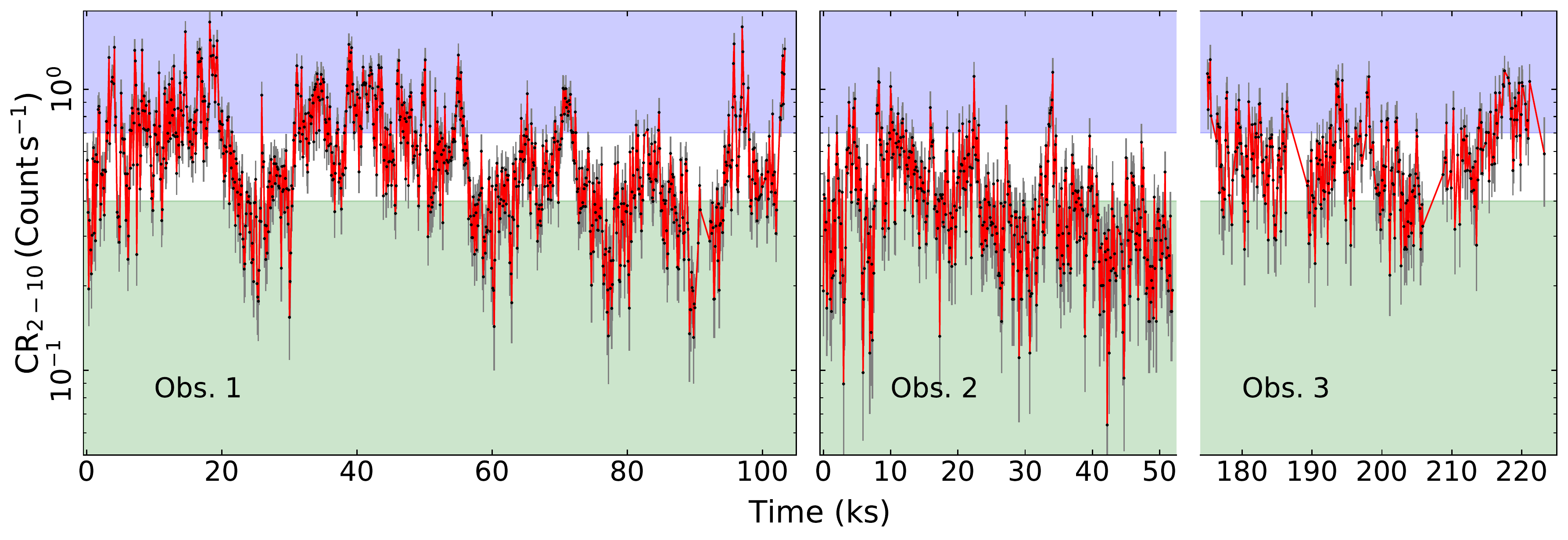}

\caption{Background-subtracted \xmm\ light curves in the $0.5-2$~keV (top panel) and $2-10$~keV (bottom panel) bands, with a time bin of 100~s. The time in Obs. 3 is from the start of Obs. 2. The green, white and blue areas correspond to the low-, medium- and high-flux levels considered in this work (see Section \ref{sec:spectral_fits} for details).}

\label{fig:xmm_LC}
\end{figure*}

X-ray observations of NGC~4395 revealed strong variability in the soft X-rays (below $\sim 2$~keV)  attributed to a complex multizone ionized absorber \citep[e.g.,][]{Iwasawa00, Dewangan08}. \cite{Nardini11} (hereafter NR11) studied the time-resolved spectra obtained from two long observations with \xmm\ and {\it Suzaku} with the aim of explaining the anomalously flat X-ray spectrum of NGC~4395 \citep[see e.g.,][]{Moran05}. They found that the source exhibited partial occultation by cold material with column densities $N_{\rm H} \sim 10^{22}-10^{23}~\rm cm^{-2}$, consistent with a clumpy broad-line region. These results were later confirmed by \cite{Parker15} who studied the X-ray variability of this source applying a `Principle Component Analysis' (PCA) to the \xmm\ data. Their results show that the variability in NGC 4395 is accounted for by a combination of intrinsic flux variability and changes in the absorption covering fraction, with hints of changes also in the column density of the absorbing material. The X-ray spectra of this source show a prominent narrow Fe K line, attributed to neutral reflection by distant material \citep[e.g.,][NR11]{Iwasawa00, Iwasawa10}. This is consistent with the PCA that did not show any hint of reflection variability. This was also confirmed by \cite{Kara16} who did not detect any evidence of low-frequency hard lag or Fe K lags from all three \xmm\ observations, despite the fact that \cite{Demarco13} have detected hints of soft lags using the first \xmm\ observation only. We note that these lags could potentially be produced by reprocessing from the warm absorber(s) as shown by \cite{Silva16} for NGC~4051.

In this work, we analyze the flux-resolved spectra of three \xmm\ observations (including the one studied by NR11) in addition to a short \nustar\ observation. Observations and data reduction are described in Section~\ref{sec:reduction}. The spectral analysis is presented in Section \ref{sec:spectral_fits}. In Section \ref{sec:timing} we show the fractional rms variability obtained from the different observations, in addition to the power spectral density using the \xmm\ observations. Finally, we discuss the results in Section \ref{sec:discussion} and we present our conclusions in Section \ref{sec:conclusion}.
\section{ Observations and data reduction}
\label{sec:reduction}

\subsection{{\it XMM-Newton} observations}
\label{sec:xmm}

NGC~4395 was observed by \xmm\ \citep{Jans01} on 2003-11-30 (Obs ID 0142830101, hereafter Obs. 1), for a total duration of $\sim 113~ \rm ks$. The time-resolved spectra from this observation have been presented by NR11. The source was later observed on 2014-12-28 and 30 (Obs IDs 0744010101 and 0744010201, hereafter Obs. 2+3, respectively) for a duration of $\sim 53$~ks each. We reduced the data from the three observations using {\tt SAS v.17.0.0} \citep{Gabriel04} and the latest calibration files. We followed the standard procedure for reducing the data of the EPIC-pn \citep{Stru01} CCD camera, operating in full frame mode for Obs. 1 and small window mode for Obs. 2+3, with medium filter. The data were processed using EPPROC. Source spectra and light curves were extracted from a circular region of a radius of 30\arcsec\ for all observations. The corresponding background spectra and light curves were extracted from an off-source circular region located on the same chip, with a radius approximately twice that of the source. After filtering out periods with strong background flares, the net exposure times dropped to 88.7~ks, 36.2~ks and 22.5~ks, for Obs. 1, 2+3, respectively. 

The light curves were produced using the {\tt SAS} task {\tt EPICLCCORR}. Fig.~\ref{fig:xmm_LC} shows the $0.5-2$~keV and $2-10$~keV light curves for the three observations, clearly revealing the large variability of this source. We note that in Obs. 1 the source was significantly brighter and more variable in the $0.5-2$~keV range compared to Obs. 2+3, while it  shows a similar brightness and variability amplitude for all observations in the $2-10$~keV band. Given the consistency between Obs. 2 and Obs. 3, and in order to increase the signal-to-noise ratio we combine the two observations (hereafter Obs. 2+3) in the rest of this work.

\subsection{{\it NuSTAR} observations}
\label{sec:nustar}

\nustar\ \citep{Harrison13} observed NGC~4395 for a net exposure of 19~ks on 2013-05-10. The data were reduced utilizing the standard pipeline in the \nustar\ Data Analysis Software ({\tt NUSTARDAS}\,v1.8.0), and using the latest calibration files. We cleaned the unfiltered event files with the standard depth correction. We reprocessed the data using the ${\tt saamode = optimized}$ and ${\tt tentacle = yes}$ criteria for a more conservative treatment of the high background levels in the proximity of the South Atlantic Anomaly. We extracted the source and background light curves and spectra from circular regions of radii 60\arcsec\ and 120\arcsec, respectively, for both focal plane modules (FPMA and FPMB) using the {\tt HEASOFT} task {\tt nuproducts}. 

\begin{figure}
\centering

\includegraphics[width = 1.\linewidth]{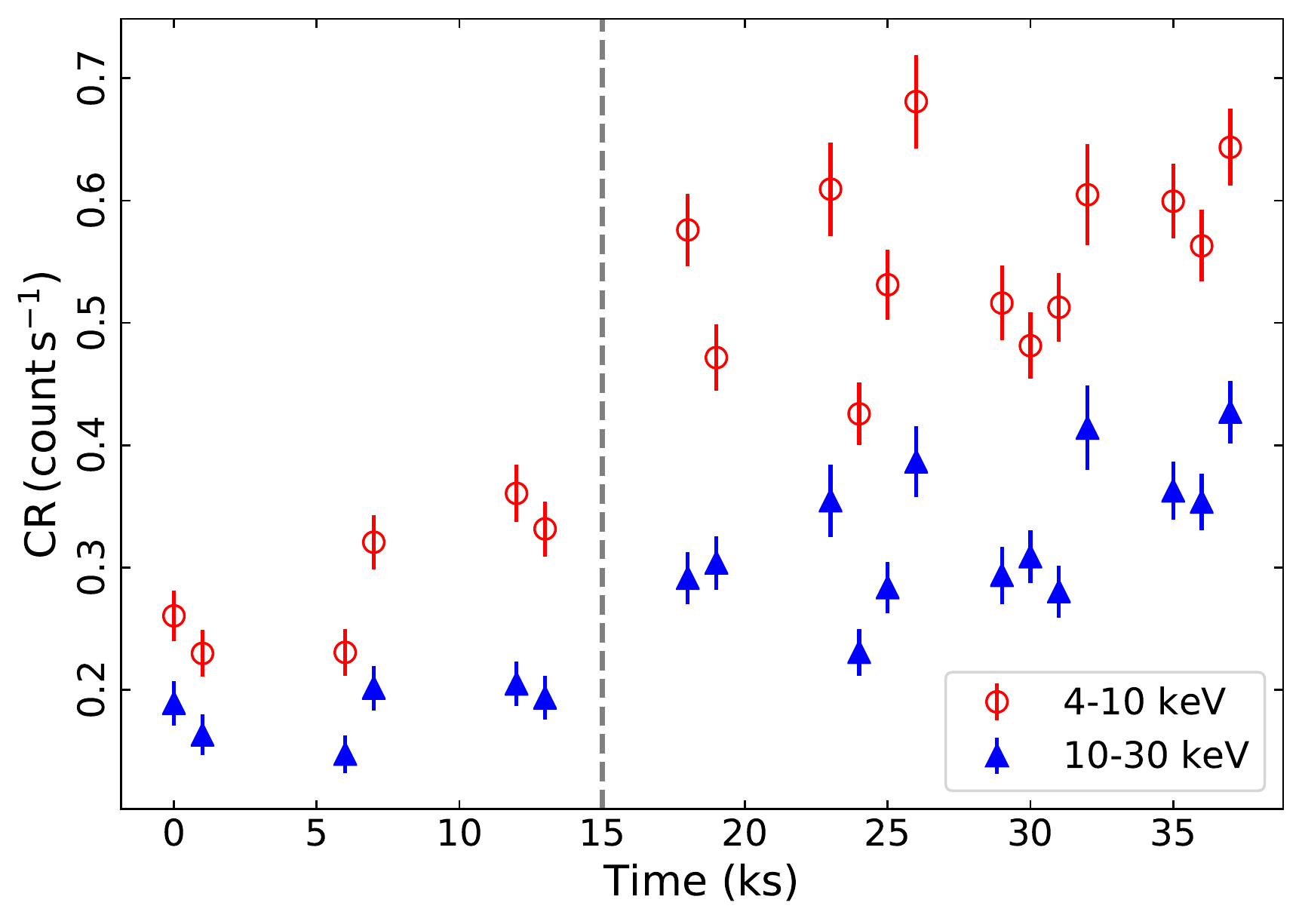}

\caption{Background-subtracted \nustar\ light curves in the $4-10$~keV (open circles) and $10-30$~keV (filled triangles) bands, with a time bin of 1~ks and an exposure fraction $> 0.5$. The light curves of FPMA and FPMB are added together. The medium- and high-flux spectra are extracted from times below/above the flux increase at $\sim 15$~ks (dashed vertical line; see see Section~\ref{sec:spectral_fits} for details).}
\label{fig:nustar_LC}
\end{figure}

We added the FPMA and FPMB light curves, using the {\tt FTOOLS} \citep{Blackburn95} command {\tt LCMATH}. Figure~\ref{fig:nustar_LC} shows the background-subtracted light curves in the $4-10$~keV and $10-30$~keV bands with a time bin of 1~ks. The source varies simultaneously in both energy bands on the timescales probed by the observations.

\section{Timing analysis}
\label{sec:timing}
In this section we investigate the variability seen in NGC 4395 through two model-independent approaches. First, we present the fractional rms variability amplitude ($F_{\rm var}$) for the various observations. Then we analyze the Power Spectral Density (PSD) for the various \xmm\ observations.

\subsection{Fractional variability}
\label{sec:Fvar}

\begin{figure}
\centering
\includegraphics[width = \linewidth]{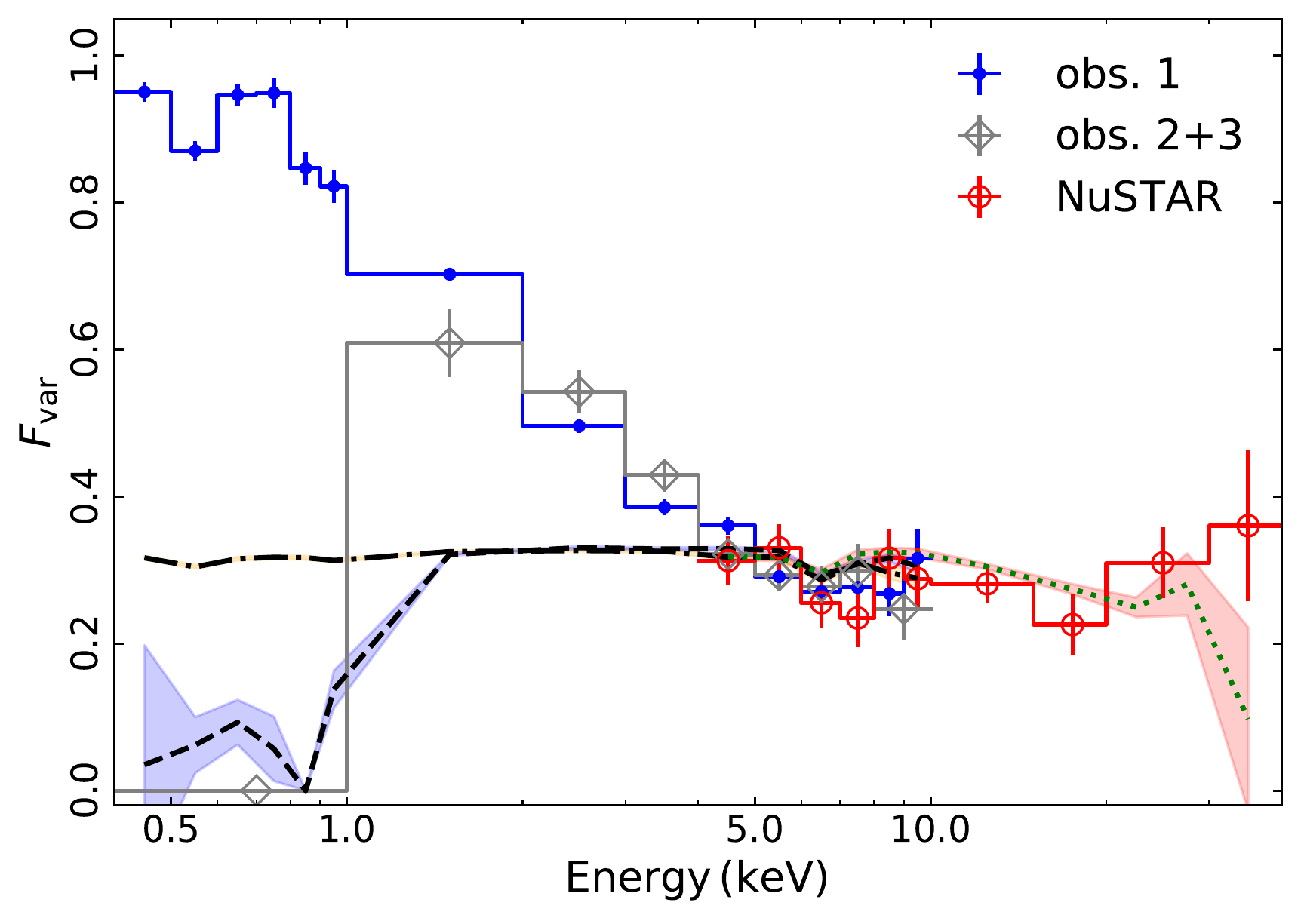}
\caption{The fractional rms variability amplitude ($F_{\rm var}$) as function of energy, obtained from the light curves, with a time bin of 1~ks, of Obs. 1 (blue dots), Obs. 2+3 (grey diamonds) and the \nustar\ observation (open red circles). The black dash-dotted  line corresponds to simulated $F_{\rm var}$, using \xmm\ responses, assuming an unabsorbed power-law varying in normalization only. The black dashed line and the green dotted line correspond to simulated $F_{\rm var}$, using \xmm\ and \nustar\ responses, respectively, assuming an absorbed power-law varying in normalization only. The column density and covering fraction of the absorber are constant and consistent with the best-fit values obtained from Obs. 2+3. The shaded areas correspond to the 1$\sigma$ uncertainty on the simulated $F_{\rm var}$ (see Section \ref{sec:discussion} for more details about the simulations). The simulated $F_{\rm var}$ are in agreement with the observations above 4~keV, while they could not reproduce the observed variability at softer energies.}
\label{fig:Fvar}
\end{figure}

It is clear from Figs.~\ref{fig:xmm_LC} and \ref{fig:nustar_LC} that the source is highly variable in all energy ranges. We characterize the variability by estimating $F_{\rm var}$ and its corresponding error, following \cite{Vaughan03}. We estimate $F_{\rm var}$ from the 1-ks binned light curves of all observations in various energy bands as shown in Fig.~\ref{fig:Fvar}. The segment lengths that are used to estimate $F_{\rm var}$ are 103~ks, 97~ks, and 37~ks for Obs. 1, 2+3, and \nustar, respectively, corresponding to the full observation lengths, after filtering. We note that due to the low flux  and the low variability level below $\sim 1$~keV in Obs. 2+3, compared with Obs. 1, it was not possible to estimate $F_{\rm var}$ in small energy bins. For that reason, we limit our analysis to one bin in the 0.4-1~keV range.

Fig. \ref{fig:Fvar} shows that $F_{\rm var}$ is constant ($\langle F_{\rm var} \rangle \simeq 0.28$) above 5~keV, for all observations. This indicates that the variability in this range is mainly due to variations in the intrinsic flux of the nuclear emission (assumed to have a power law shape varying in normalization only), in each observation. Some deviations, though not statistically significant, can be seen in the $6-7$~keV and $15-20$~keV ranges, where the relative contribution of reflection features (Fe K emission line and Compton hump, respectively) is expected to be larger. Longer \nustar\ observations would be needed to confirm the presence of these features, especially above 10~keV. However, $F_{\rm var}$ increases below $\sim 4$~keV in Obs. 1, reaching more than $\sim 0.9$. This could be mainly due to the complex variable absorption structure in this source affecting mainly the soft X-rays (see NR11 and Section \ref{sec:spectral_fits} in the current work, for more details about the spectral modeling). Intrinsic variability due to a change in the flux of the power-law component would result in a nearly constant $F_{\rm var}$ at all energies. However, any additional variability process that might be caused by a change in absorption, for example, would lead to an increase in the values of $F_{\rm var}$ in the energy range affected by these changes \citep[see e.g.,][ and Section~\ref{sec:discussion} for more details]{Matzeu16}. We note that that this does not necessarily imply that the intrinsic continuum fluctuations and the absorption changes contributing to the variability in the soft X-rays are operating on the same timescales. Any changes in the absorber column density and/or the covering fraction are expected to occur on longer timescales than the continuum. Hence, what drives the increase in $F_{\rm var}$ is the large amplitude (a factor of more than 30 compared to $\sim 3$ below and above 2~keV, respectively, as seen in Fig.~\ref{fig:xmm_LC}) of the variations associated with such changes. As for Obs. 2+3, $F_{\rm var}$ is zero in the $0.4-1$~keV indicating a low variability amplitude in this energy range.

\subsection{The power spectral density}
\label{sec:psd}

\begin{figure*}
\centering

\includegraphics[width = 1.\linewidth]{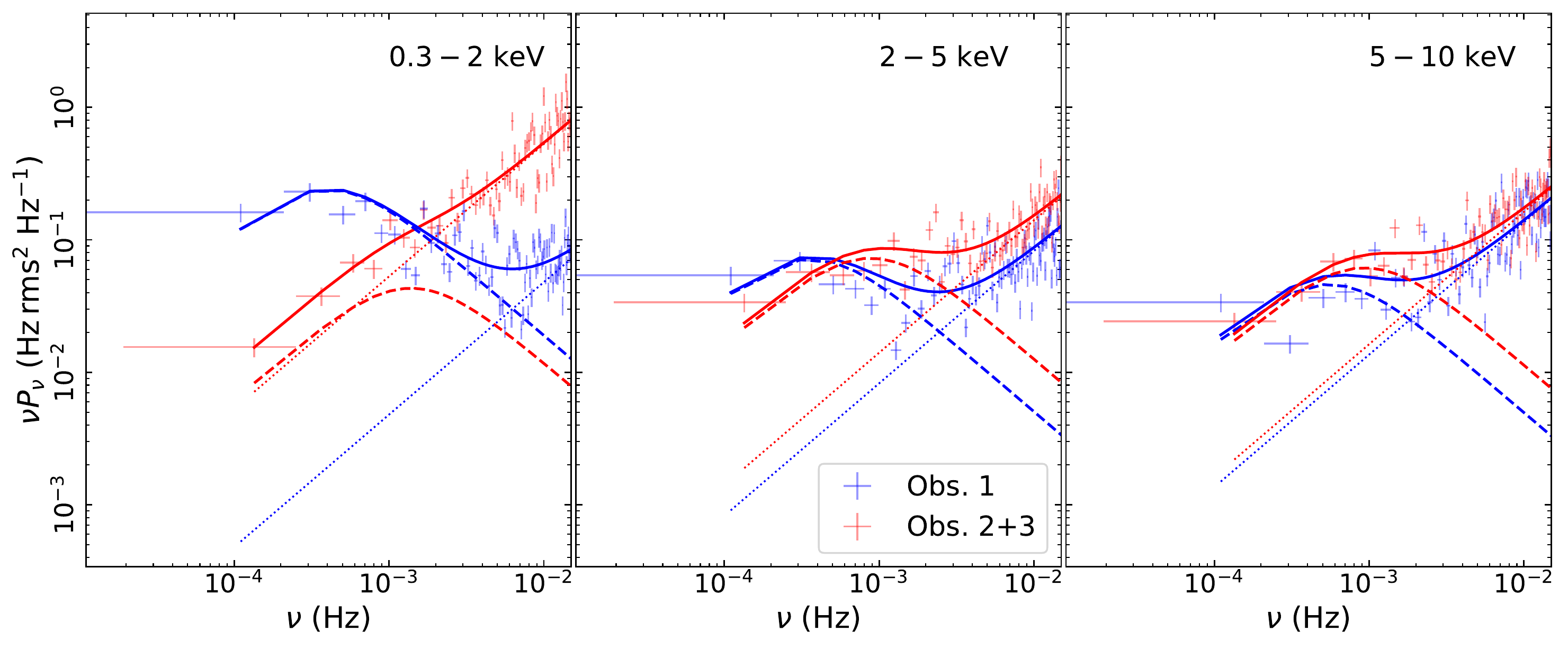}\\
\includegraphics[width = 1.\linewidth]{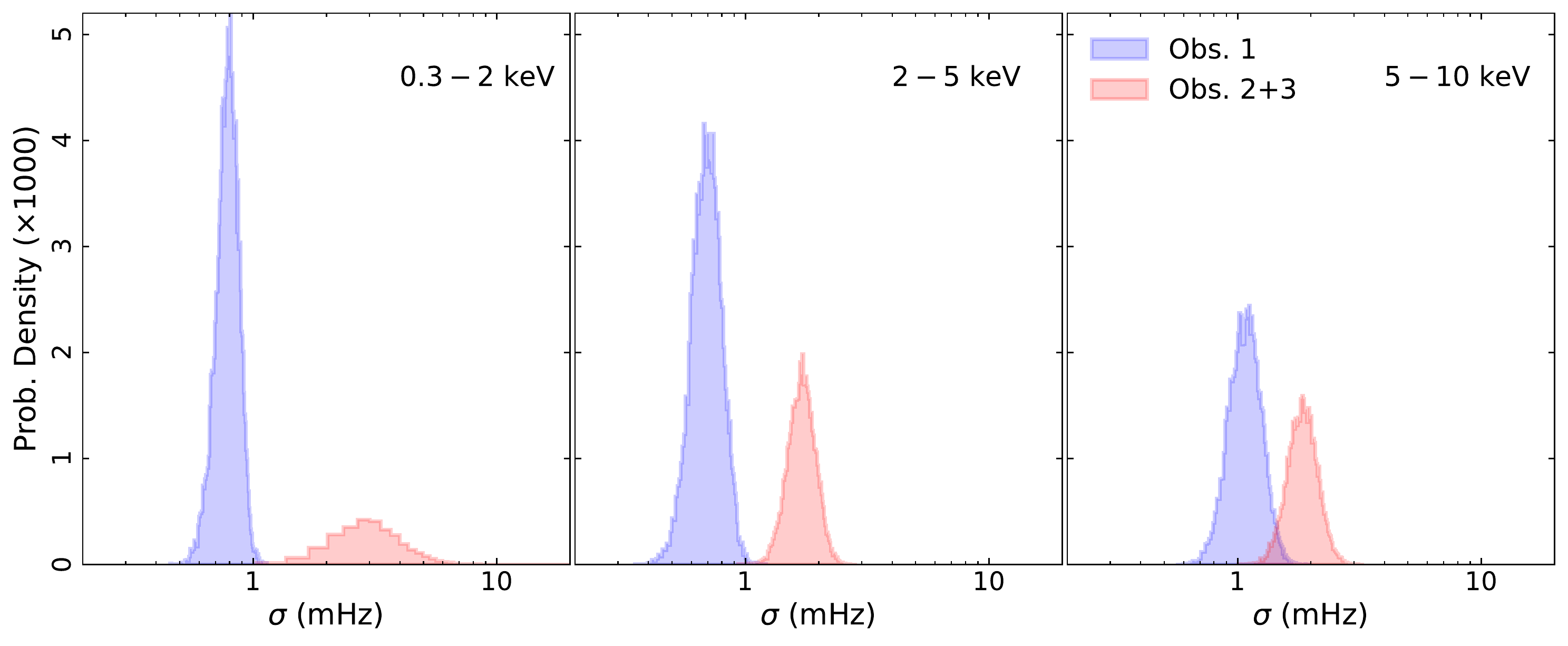}
\caption{Top panel: Power spectral density for Obs. 1 (blue) and Obs. 2+3 (red) in the $0.3-2$~keV, $2-5$~keV, and $2-10$~keV bands. The solid lines show the best-fit models, the dashed lines correspond to the best-fit Lorentzian, the dotted lines represent the Poisson noise level. Bottom panel: The probability density of double the characteristic frequency ($\sigma = 2\nu_{\rm peak}$) for each case, obtained from the MCMC analysis (see Section \ref{sec:psd} for details).}
\label{fig:PSD}
\end{figure*}

As an additional model-independent variability study, we estimate the PSD of the source in different energy bands, using the periodogram \citep[e.g.,][]{Vaughan03}. Background-subtracted light curves with a sampling time of 16~s are extracted from standard event files as described in Section \ref{sec:xmm}. Gaps smaller than 400~s resulting from the standard background filtering are linearly interpolated and randomized. The periodogram using the RMS normalization \citep[see][]{Vaughan03} is then calculated from the Discrete Fourier transform of the light curve arrays. The final periodogram is produced by averaging every 20 frequency values. The modeling is done in {\tt XSPEC} using the Whittle20 statistic \citep{Whittle53, Vaughan10, Barret12}. A simple power law does not account for the apparent break at $ \sim$ a few $\times 10^{-4}$~Hz \citep[see][]{Vaughan05}. Hence, to estimate the characterstic time scale, the periodograms are modeled with a zero centered Lorentzian of the form,
\begin{equation}
P(\nu) = K \frac{\sigma}{2\pi} \frac{1}{\nu^2 + (\sigma/2)^2} ,
\end{equation}
where $K$ is the normalization. In the $\nu P(\nu)$ representation, the peak frequency $\nu_{\rm peak} = \sigma/2$ can be considered as the ``characteristic frequency'' of the PSD \citep[see e.g.,][]{Belloni97, Belloni02}. The Poisson noise is modeled with a constant for each PSD. The errors on the model parameters are estimated using the  Goodman-Weare MCMC algorithm in {\tt XSPEC} \citep{Goodman10}. We estimate the PSDs separately from Obs. 1 and Obs. 2+3 (combined together). The top panel of Fig. \ref{fig:PSD} shows the PSD of NGC 4395 in the $0.3-2$~keV, $2-5$~keV and $5-10$~keV bands. The best-fit $\sigma$  and normalization of the Lorentzian are listed in Table \ref{table:psd}.

\begin{table}
\centering
\caption{The best-fit $\sigma$ and normalization obtained by fitting the PSDs with a Lorentzian function in the $0.3-2$~keV, $2-5$~keV, and $2-10$~keV bands.}
\begin{tabular}{lll}
\hline \hline
	&	Obs. 1	&	Obs. 2+3	\\ \hline
	&	$0.5 - 2$~keV	&		\\ 
$\sigma~ \rm (mHz)$	&$	0.79\pm 0.08	$&$	3.43^{+0.57}_{-1.40}	$ \\ \\[-0.2cm]
Norm $(\times 10^{-1})$	&$	7.59^{+0.51}_{-0.81}	$&$	1.48^{+0.23}_{-0.35}	$ \\ \\[-0.2cm] \hline
	&	$2 - 5$~keV	&		\\
$\sigma~ \rm (mHz)$	&$	0.71^{+0.09}_{-0.11}	$&$	1.74^{+0.21}_{-0.24}	$ \\ \\[-0.2cm]
Norm $(\times 10^{-1})$		&$	2.26^{+0.19}_{-0.26}	$&$	2.30^{+0.16}_{-0.20}	$ \\ \\[-0.2cm] \hline
	&		$5 - 10$~keV&		\\ 
$\sigma~ \rm (mHz)$	&$	1.11^{+0.15}_{-0.18}	$&$	1.90^{+0.23}_{-0.30}	$ \\ \\[-0.2cm]
Norm $(\times 10^{-1})$		&$	1.46^{+0.11}_{-0.15}	$&$	1.96^{+0.15}_{-0.17}	$ \\ \\[-0.2cm] \hline \hline
\end{tabular}
\label{table:psd}
\end{table}


The probability densities of $\sigma$ for each case are shown in the lower panel of Fig. \ref{fig:PSD}. This figure shows that the characteristic frequency is smaller in Obs. 1 compared to Obs. 2+3, for all energy bands. Furthermore, the values of $\sigma$ in the $0.3-2$~keV and $2-5$~keV bands are consistent during Obs. 1, and smaller than the best-fit value in the $5-10$~keV band. This behavior could be due to the fact that Obs. 1 is affected by both intrinsic variability and an additional variability process, most probably absorption changes (as discussed in NR11 and Section \ref{sec:fit} of the current work), though operating on different timescales. The additional variability process, also seen in the $F_{\rm var}$, is expected to occur on longer timescales compared to the intrinsic variability, and to affect mainly the soft X-rays. Hence, its contribution to the overall variability decreases as the energy increases, which might explain the shift to shorter timescales in Obs. 1 as the intrinsic variability becomes more dominant (in the hard X-rays). However, in Obs. 2+3, the intrinsic variability seems to be dominating over all energy bands, which could explain the fact that the values of $\sigma$ are consistent over the full $0.3-10$~keV range. We note that the apparent large value of $\sigma$ in the $0.3-2$~keV band in Obs. 2+3 is mainly due to the data quality as signal is heavily affected by the Poisson noise. The value of $\sigma$ in the $5-10$~keV range in Obs. 1 is systematically smaller than the one in Obs. 2+3 (they are consistent within $\sim 2.5\sigma$). This is most likely due to the relatively high column density of the variable neutral absorber in Obs. 1 (as discussed in Section \ref{sec:fit}) which still affects the overall variability in this energy range (though in a more moderate way), shifting the characteristic timescale towards a slightly larger value.

It is known that there is a relation between the power-spectrum break timescale, BH mass, and X-ray luminosity established for unobscured AGN and Galactic BH systems, of the form,

\begin{equation}\label{eq:McHardy}
\log  T_{\rm B}  = 2.1 \log(M_{\rm BH,6}) - 0.98 \log(L_{\rm bol, 44}) - 2.32,
\end{equation}
\begin{equation}\label{eq:GonzalezMartin}
\log  T_{\rm B}  = 1.39 \log(M_{\rm BH,6}) - 0.82 \log(L_{\rm bol, 44}) - 2.7,
\end{equation}
\noindent as derived by \cite{McHardy06} and \cite{GonzalezMartin18}, respectively. In both relationships, $T_{\rm B}$ is the break timescale in units of day, $M_{\rm BH,6}$ is the BH mass in units of $10^6~M_\odot$ and $L_{\rm bol,44}$ is the bolometric luminosity in units of $10^{44}~\rm erg~s^{-1}$. We note that in our modeling we do not assume a broken shape of the PSD. If these relationships hold also for NGC 4395, then considering $T_{\rm B} = \nu_{\rm peak}^{-1} = (\sigma/2)^{-1}$, with  $\sigma = 1.9~\rm mHz$ (considering mainly the intrinsic variability above 5~keV in Obs. 2+3; see Table 5), and a bolometric lumninosity\footnote{\cite{Lira99} estimated $L_{\rm bol} = 1.2\times 10^{41}~\rm erg~s^{-1}$, for $d=5.2~\rm Mpc$ and considering the SED below 2~keV. However, for $d=4.2~\rm Mpc$ \citep{Karachentsev98}. Adding the $2-100$~keV luminosity assuming the best-fit from the medium-flux \nustar\ spectrum (see Section \ref{sec:fit}) we get $L_{\rm bol}= 1.91\times 10^{41}~\rm erg~s^{-1}$.} of $ 1.52\times 10^{41}~\rm erg~s^{-1}$, we infer $M_{\rm BH} = 8.4 \times 10^4~\rm M_\odot$ and $9.2 \times 10^4~\rm M_\odot$, following eq. \ref{eq:McHardy} and \ref{eq:GonzalezMartin}, respectively.  \cite{GonzalezMartin18} showed also that obscuration events might affect the $ T_{\rm B} - M_{\rm BH}$ relationship. Our mass estimate is intermediate compared to the previously reported estimates of $9.1_{-1.6}^{+1.5} \times 10^3~\rm M_\odot$ \citep{Woo19} and $(3.6 \pm 1.1)\times 10^5~\rm M_\odot$ \citep{Peterson05}. It is worth mentioning that this bolometric luminosity would correspond to and Eddington ratio $L_{\rm bol}/L_{\rm Edd} = 0.12~(0.003)$ for $M_{\rm BH} = 10^4 ~ (3.6\times 10^5)~\rm M_\odot$.

\section{X-ray Spectral analysis}
\label{sec:spectral_fits}

\begin{table}
\centering
\caption{Net count rate (in the $0.5-3$~keV and $3-10$~keV bands) and net exposure time for each flux-resolved spectrum obtained from the different observations. Count rates are in units of $\rm count~s^{-1}$, exposure times are in units of ks.}
\begin{tabular}{llll}
\hline \hline
	&		Low		&		Mean		&		High		\\ \hline
Obs. 1	&				&				&				\\
$\rm CR_{0.5-3}$	&	$	0.202 \pm 0.003	$	&	$	0.435 \pm 0.004	$	&	$	1.131 \pm 0.007	$	\\
$\rm CR_{3-10}$	&	$	0.199 \pm 0.003	$	&	$	0.331 \pm 0.003	$	&	$	0.55 \pm 0.005	$	\\
Net exposure	&		23.3		&		38.9		&		26.5		\\ \hline
Osb. 2+3	&				&				&				\\
$\rm CR_{0.5-3}$	&	$	0.076 \pm 0.002	$	&	$	0.113 \pm 0.002	$	&	$	0.199 \pm 0.005	$	\\
$\rm CR_{3-10}$	&	$	0.215 \pm 0.003	$	&	$	0.384 \pm 0.004	$	&	$	0.598 \pm 0.009	$	\\
Net exposure 	&		24.8		&		25.2		&		8.7		\\ \hline
\nustar/FPMA	&				&				&				\\
$\rm CR_{3-10}$	&	$	-	$	&	$	0.126 \pm 0.004	$	&	$	0.218 \pm 0.004	$	\\
Net exposure (ks)	&	$	-	$	&	$	6.9	$	&	$	12	$	\\ \hline
\nustar/FPMB	&				&				&				\\
$\rm CR_{3-10}$	&	$	-	$	&	$	0.113 \pm 0.004	$	&	$	0.224 \pm 0.004	$	\\
Net exposure	&	$	-	$	&	$	6.9	$	&	$	12	$	\\ \hline \hline

\end{tabular}
\label{table:Counts}
\end{table}

Following a complementary approach with respect to the time-resolved analysis of NR11, here we consider the flux-resolved spectra from all observations. For \xmm, we extract the spectra for three different flux levels, using the EPIC-pn light curves: low, medium and high, with $2-10$~keV count rate below $\rm 0.4~ count~s^{-1}$, between  $\rm 0.4~ count~s^{-1}$ and  $\rm 0.7~ count~s^{-1}$, and above  $\rm 0.7~ count~s^{-1}$, respectively (as shown in Fig.~\ref{fig:xmm_LC}). The \xmm\ spectra from the different flux levels and observations are grouped requiring a minimum signal-to-noise ratio (S/N) of 4 per energy bin. We have checked also the EPIC-MOS data. The data from Obs. 3 are heavily affected by background flares and cannot be used. Moreover, matching the count rates between the various detectors to define the flux states is not trivial, and adding the MOS spectra would be of a little help in the low flux states due to the poor data quality, and redundant for the high flux flux levels. Hence, for simplicity, we decided to neglect the MOS data. We also checked the RGS data. However, they are dominated by noise, even when considering the time-averaged spectra. For that reason we do not include them in this analysis. 

\begin{figure*}
\centering

\includegraphics[width = 0.99\linewidth]{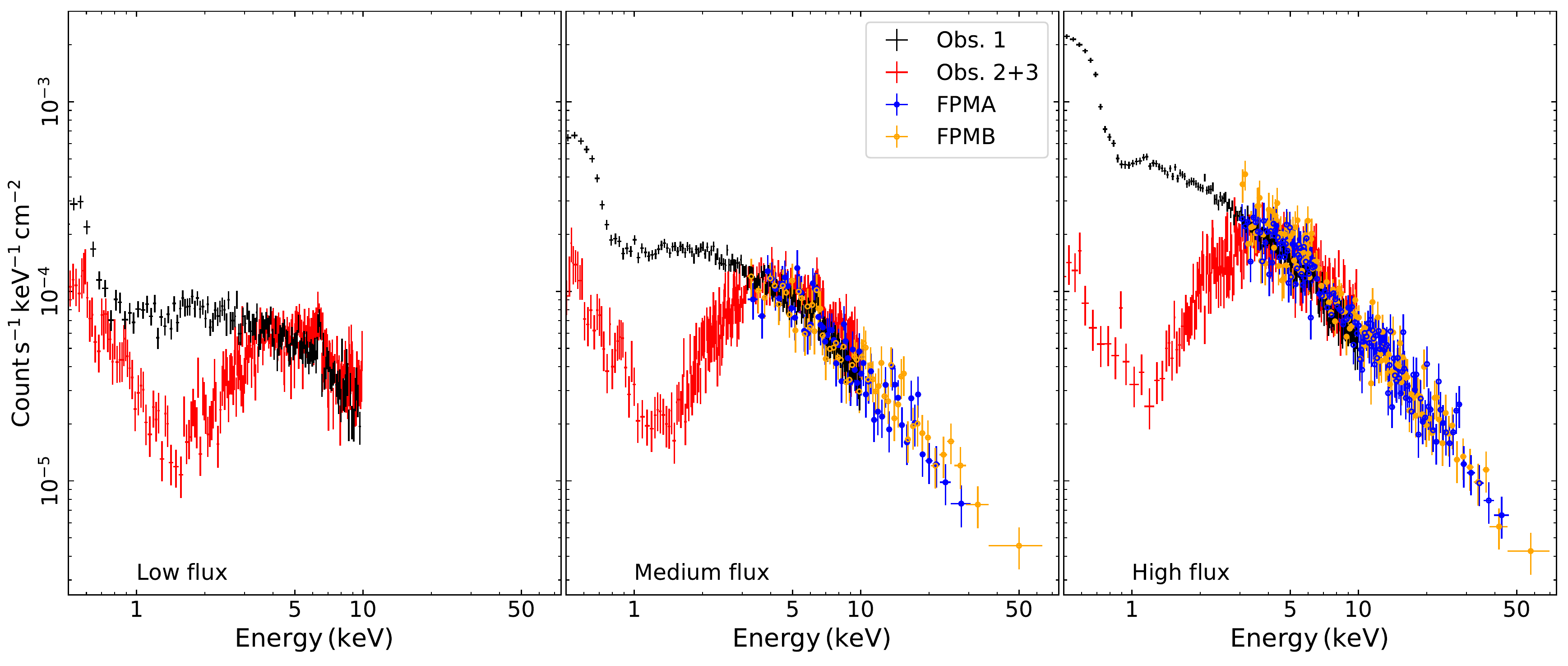}

\caption{Low-, medium- and high-flux spectra (left, middle, and right panel) from Obs. 1 (black), Obs. 2+3 (red), and the \nustar\ observation (FPMA and FPMB in blue and orange dots, respectively).}
\label{fig:spectra_FR}
\end{figure*}

The \nustar\ light curves presented in Fig.~\ref{fig:nustar_LC} show an increase in flux after $\sim 15$~ks from the start of the observation. Thus, we extract spectra before and after the flux increase at 15~ks, requiring a minimum S/N of 4 per energy bin. The spectra extracted from FPMA and FPMB (extending up to 70~keV) are consistent with each other, for both flux levels, so they are analyzed jointly (but not combined together). Table~\ref{table:Counts} shows the net count rate (in the $0.5-3$~keV and $3-10$~keV bands) and the net exposure time for each flux level in the different observations. The spectra from all flux levels are shown in Fig.~\ref{fig:spectra_FR}. It can be seen that the soft X-rays (below $\sim 1$~keV) vary by a factor of more than $\sim 6$ in Obs. 1 (between the low- and high-flux levels) while they are consistent with each other for all flux levels in Obs. 2+3. The spectra from all flux levels in Obs. 1, 2+3 are consistent above $\sim 4$~keV. As for the \nustar\ observation, the spectra extracted from the first part of the observation are consistent with the medium-flux spectra extracted from the \xmm\ observations, and the spectra from the second part of the observation are consistent with the high-flux \xmm\ spectra.

Throughout this work, spectral fitting is performed using XSPEC\,v12.10e \citep{Arnaud96}. The \xmm\ spectra are fitted in the $0.5-10$\,keV range, while the \nustar\ spectra are fitted in the $3-70$~keV range. We apply the $\chi^2$ statistic using the ``model'' weighting. This weighting method estimates errors on each bin based on the model-predicted number of counts rather than the square root of the number of counts, which can introduce a bias in the fit to low flux states at modest count rates. We list the uncertainties on the parameters at the $90\%$ confidence level ($\Delta \chi^2= 2.71$), unless stated otherwise. These uncertainties are calculated from a Markov chain Monte Carlo (MCMC)\footnote{We use the {\tt XSPEC\_EMCEE} implementation of the {\tt PYTHON EMCEE} package for X-ray spectral fitting in {\tt XSPEC} by Jeremy Sanders
(\url{http://github.com/jeremysanders/xspec\_emcee})} analysis, starting from the best-fitting model that we obtained. We used the Goodman-Weare algorithm \citep{Goodman10} with a chain of $5\times 10^5$ elements, discarding the first 30\% of elements as part of the ‘burn-in’ period. 

\subsection{Spectral fitting}
\label{sec:fit}

\begin{figure}
\centering

\includegraphics[width = 1.\linewidth]{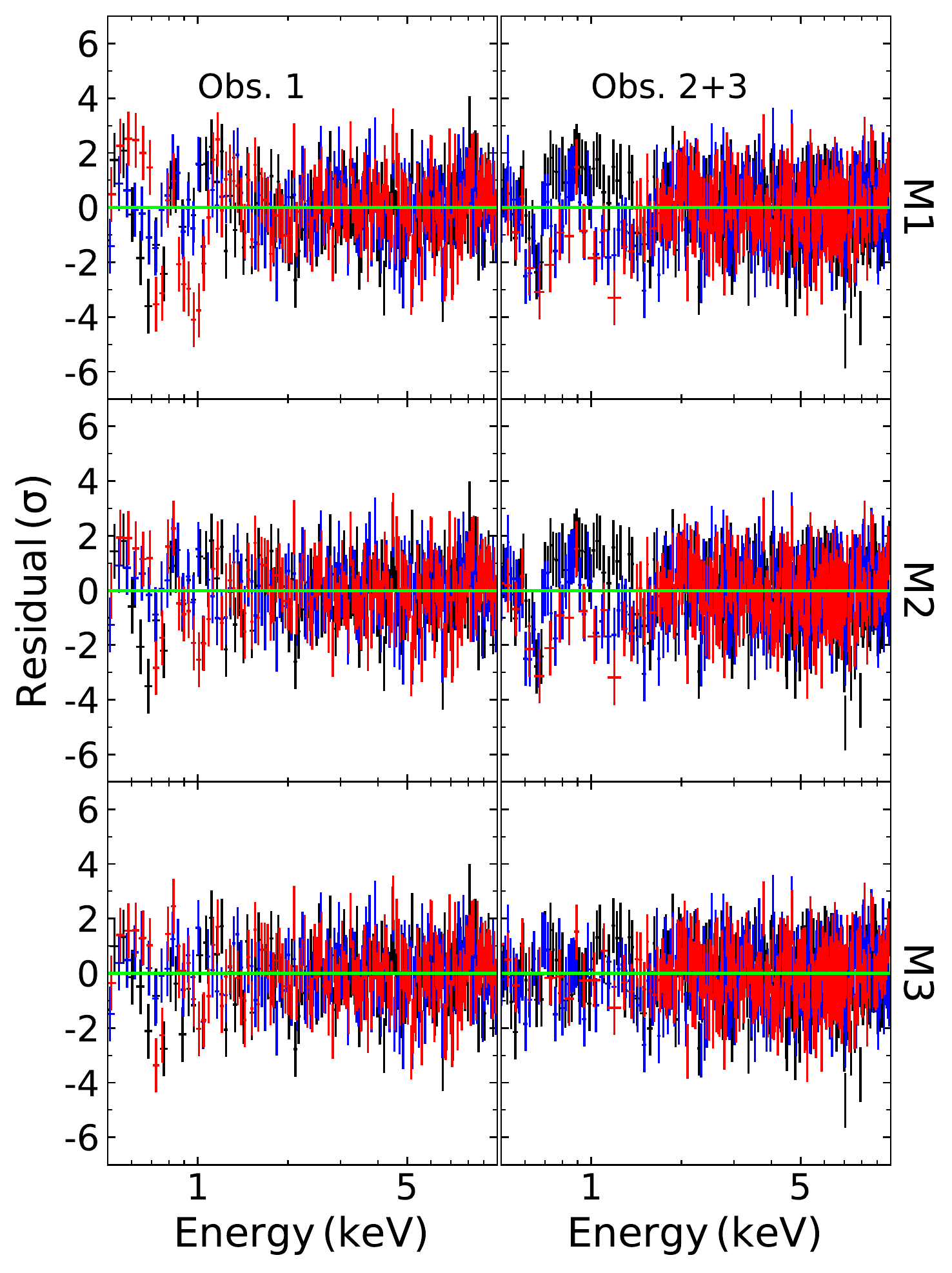}

\caption{Residuals for the different models (${\tt M1-M3}$, top to bottom) obtained by adding various absorption/emission components (see text for details), for each \xmm\ observation. The black, blue and red spectra in each panel correspond to the low-, medium- and high-flux spectra, respectively. The \nustar\ residuals are not shown because the added spectral components affect mainly the soft energies (below $\sim 2$~keV).}
\label{fig:spectra_residuals}
\end{figure}

Previous studies \citep[e.g., ][NR11]{Iwasawa00, Dewangan08, Iwasawa10} show that the X-ray spectrum of NGC~4395 is composed of a primary power-law (PL) continuum that is affected by complex neutral and ionized absorption, in addition to a neutral reflection component. Moreover, the {\it Chandra} and the {\it HST} [\ion{O}{3}] images of the source \citep[see e.g.,][]{Gomez17} show an extended soft emission region. Hence, we fitted the spectra using a PL model with a high energy cutoff, modified by neutral and ionized partial covering absorbers. We also added a neutral reflection component and an emission component from a collisionally ionized diffuse gas representing the contribution from the extended regions. The model is written in XSPEC terminology as follows:

\begin{eqnarray}
\begin{array}{lc}
{\tt M1} & = {\tt phabs[1] * (zpcfabs[2] * zxipcf[3] * zcutoffpl[4] }
\end{array}\nonumber\\
\begin{array}{ll}
 {\tt + pexmon[5] + apec[6]),} & 
\end{array}\nonumber
\end{eqnarray}

\noindent where ${\tt phabs[1] }$ accounts for Galactic absorption in the line of sight (LOS) of the source \citep[$N_{\rm H} = 4.34 \times 10^{20}~\rm cm^{-2}$;][]{HI4PI} and ${\tt zpcfabs[2], zxipcf[3] }$ \citep{Reeves08} account for the neutral and ionized absorption, respectively, at the redshift of the source. Neutral reflection is modeled using ${\tt pexmon[5]}$\footnote{The data do not require any ionized reflection. Hence, using a different model for the reflection, such as {\tt Xillver} \citep{Garcia10, Garcia13}  which includes emission lines from more elements as compared to {\tt pexmon}, would not affect our results, as the contribution of these emission lines to the soft spectrum would be negligible for neutral reflection.} \citep{Nandra07} and diffuse emission is modeled using ${\tt apec[6]}$ \citep{Smith01}. 

We kept the photon index of {\tt cutoffpl} as a free parameter for Obs. 1 but tied among the three flux levels. For the rest of the observations (Obs. 2+3 and \nustar), we kept $\Gamma$ tied to a single free value jointly determined by the flux-selected spectra. We fixed the high-energy cutoff to 500~keV and let the normalization be free for all the spectra. For the {\tt pexmon} model we linked the photon index and high-energy cutoff to the corresponding {\tt cutoffpl} parameters. We fixed the abundance to the solar value and the inclination to 45\degr. The normalization of {\tt pexmon} is free for Obs. 1 and Obs. 2+3 (tied between the three flux levels) \footnote{We test the assumption of having a constant reflection component by fitting the spectra for each observation separately in the $5.5-8$~keV range assuming a power law plus a Gaussian emission line. The {\tt pexmin} normalization of the \nustar\ observation is tied to the one of Obs. 2+3. We found that the flux of the line is constant within each observation for all flux levels, and  consistent between each observation.}. The temperature and the normalization of the {\tt apec} component are kept tied for all observations and flux levels. 

\begin{figure*}
\centering

\includegraphics[width = 1.\linewidth]{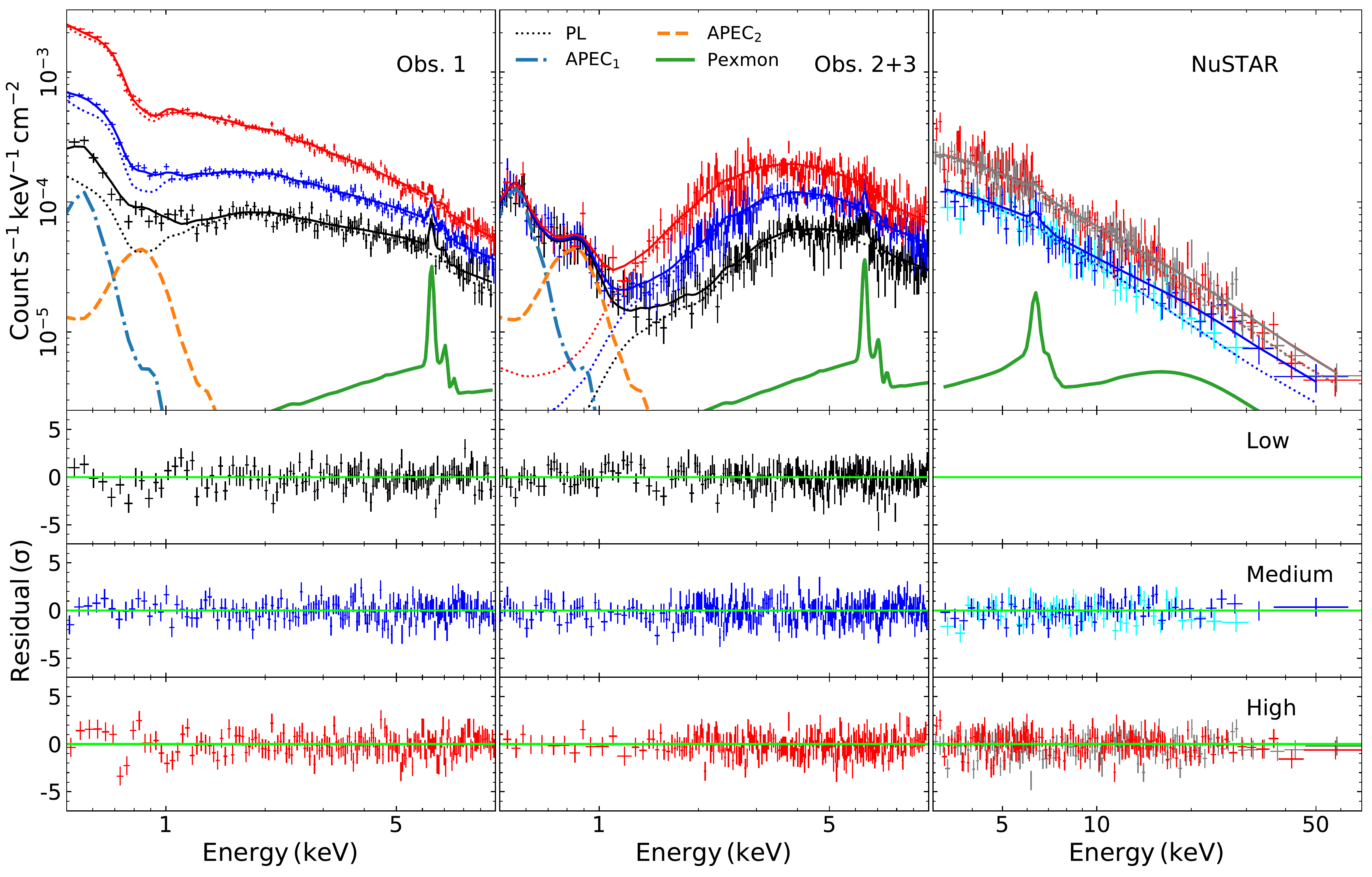}

\caption{Top panel: Spectra and best-fitting model (${\tt M3}$) for all observations. The black, blue and red spectra in each panel correspond to the low-, medium- and high-flux spectra, respectively. The blue/cyan and the red/grey \nustar\ spectra correspond to the medium- and high-flux FPMA/B spectra, respectively. The thin dotted lines represent the transmitted power-law component. The thick green ines represent the reflection component, the blue dashed-dotted lines and the orange dashed lines correspond to the soft diffuse thermal emission components.Bottom panels: corresponding residuals for each flux level.}
\label{fig:spectra_obs}
\end{figure*}

We let the column density of the neutral absorber ($N_{\rm H, N}$) and the corresponding covering fractions be free for all of the spectra. As for the \nustar\ spectra, due to the lack of a simultaneous high-quality observation in the soft X-rays, it is hard to identify and characterize all the possible absorption components. Hence, we conservatively assume neutral absorption fully covering the source. Fitting the \nustar\ data for ionized absorption resulted in unconstrained parameters and a covering fraction consistent with zero. For that reason we do not consider this component in the fit. The fit suggested also a fully covering neutral absorber for Obs. 2+3 with a constant $N_{\rm H, N}$ across all flux levels. Therefore we fixed the neutral covering fraction to 1 and tied $N_{\rm H, N}$ for all flux levels. For Obs. 1, the fit suggests a variable covering fraction among the different flux levels. As for the ionized absorber, we kept the column densities ($N_{\rm H, W1}$) and the corresponding covering fractions free to vary between observations but tied for the different flux levels. However, we left the ionization parameter\footnote{The ionization parameter is defined as $\xi = L_{\rm ion}/nr^2$, where $n$ is the electron density of the gas and $r$ is its distance from an ionizing source with
$1-1000$~Ry luminosity $L_{\rm ion}$.} ($\log \xi_{\rm W1}$) free to vary for all of the spectra. The resulting fit is not statistically acceptable ($\rm \chi^2/dof = 1775/1537$) with residuals suggesting absorption-like features in the $0.6-1$~keV range, as seen in the top panel of Fig.~\ref{fig:spectra_residuals}.

Hence, we added another ionized absorption component, at the redshift of the source, also modeled with {\tt zxipcf}, to account for a higher ionization level absorber compared to the one already probed. The model becomes, in XSPEC terminology,

\begin{eqnarray}
\begin{array}{l}
{\tt M2}  = {\tt phabs[1]* }
\end{array}\nonumber\\
\begin{array}{l}
 {\tt (zpcfabs[2] *zxipcf[3] * zxipcf[4] * zcutoffpl[5]}  
\end{array}\nonumber\\
\begin{array}{l}
 {\tt + pexmon[6] + apec[7]).}  
\end{array}\nonumber
\end{eqnarray}

\noindent For {\tt M2} we kept the column density ($N_{\rm H,W2}$), the ionization parameter ($\log \xi_{\rm W2}$) and the covering fraction ($CF_{\rm W2}$) free to vary between Obs. 1 and the other observations. We kept $N_{\rm H,W2}$ and $CF_{\rm W2}$ tied for the different flux levels corresponding to the same observations, and $\log \xi_{\rm W2}$ free to vary between the flux levels. The covering fractions for Obs. 2+3 and the \nustar\ observation were consistent with zero, so we do not consider the presence of this component for these observations. The fit is still not statistically acceptable ($\rm \chi^2/dof = 1684/1532$) but it has improved by $\Delta \chi^2 = -91$ for five more free parameters. The improvement in Obs. 1 can be clearly seen in the second row of Fig.~\ref{fig:spectra_residuals}. Some excess emission in the $~\sim 0.6-1.5$~keV range is still seen in Obs. 2+3. This component could be accounted for by the temperature gradient that is expected to be present in the diffuse gas. For that reason, and for consistency with our previous modeling, we added another {\tt apec} component that is, conservatively, assumed to be constant for all observations. The model becomes,

\begin{eqnarray}
\begin{array}{l}
{\tt M3}  = {\tt phabs[1]* }
\end{array}\nonumber\\
\begin{array}{l}
 {\tt (zpcfabs[2] *zxipcf[3] *zxipcf[4] * zcutoffpl[5]}  
\end{array}\nonumber\\
\begin{array}{l}
 {\tt + pexmon[6] + apec[7] +  apec[8]).}  
\end{array}\nonumber
\end{eqnarray}
\noindent The fit is statistically acceptable ($\chi^2/{\rm dof} = 1568/1530 , ~ p_{\rm null} =0.24$) with $\Delta \chi^2 = -116$ for two more free parameters. The residuals shown in the third row of Fig.~\ref{fig:spectra_residuals} and the bottom rows of Fig. \ref{fig:spectra_obs} show a clear improvement in all observations with no obvious residuals.

\begin{table}
\centering

\caption{Best-fit absorption parameters. Units are as follows: column densities in $10^{22}~\rm cm^{-2}$, and ionization parameters in $\rm erg~cm~s^{-1}$.}

\begin{tabular}{lccc}
\hline \hline
Parameter	&		Low		&		Medium		&		High		\\ \hline
	&		\multicolumn{3}{c}{Obs. 1}										\\
$N_{\rm H, N}$	&	$		$	&	$	34.36_{-6.83}^{+11.09}	$	&	$		$	\\
$CF_{\rm N}$	&	$	0.48_{-0.07}^{+0.04}	$	&	$	0.38_{-0.05}^{+0.04}	$	&	$	0.16 \pm 0.06	$	\\
$N_{\rm H, W1}$	&	$		$	&	$	0.82_{-0.11}^{0.17}	$	&	$		$	\\
$\log \xi_{\rm W1}$	&	$	0.14_{-0.29}^{+0.19}	$	&	$	0.67_{-0.05}^{+0.09}	$	&	$	1.28_{-0.07}^{+0.09}	$	\\
$CF_{\rm W1}$	&	$		$	&	$	0.92_{-0.05}^{+0.04}	$	&	$		$	\\
$N_{\rm H, W2}$	&	$		$	&	$	3.81_{-0.43}^{+0.79}	$	&	$		$	\\
$\log \xi_{\rm W2}$	&	$	1.99_{-0.19}^{+0.06}	$	&	$	2.08_{-0.09}^{+0.06}	$	&	$	2.28_{-0.11}^{+0.06}	$	\\
$CF_{\rm W2}$	&	$		$	&	$	0.91_{-0.08}^{+0.05}	$	&	$		$	\\ \hline
	&		\multicolumn{3}{c}{Obs. 2+3}										\\
$N_{\rm H, N}$	&	$		$	&	$	1.62_{-0.28}^{+0.53}	$	&	$		$	\\
$CF_{\rm N}$	&	$		$	&	$	1^{\rm fixed}	$	&	$		$	\\
$N_{\rm H, W1}$	&	$		$	&	$	7.83_{-0.38}^{+0.97}	$	&	$		$	\\
$\log \xi_{\rm W1}$	&	$	0.17_{-0.17}^{+0.21}	$	&	$	0.99_{-0.35}^{+0.13}	$	&	$	1.33_{-0.11}^{+0.18}	$	\\
$CF_{\rm W1}$	&	$		$	&	$	0.95_{-0.02}^{+0.01}	$	&	$		$	\\ \hline
	&		\multicolumn{3}{c}{\nustar}										\\
$N_{\rm H, N}$	&	$		$	&	$	5.95_{-0.91}^{+2.04}	$	&	$		$	\\
$CF_{\rm N}$	&	$		$	&	$	1^{\rm fixed}	$	&	$		$	\\ \hline\hline
	
\end{tabular}
\label{table:absorptionpar}
\end{table}

The best-fit model and all components are presented in Fig.~\ref{fig:spectra_obs}. We report in Tables \ref{table:absorptionpar} and \ref{table:emissionpar} the best-fit parameters for the absorption and emission components, respectively. The best fits reveal a small change in the power law photon index from $1.67_{-0.04}^{+0.02}$ in Obs. 1 to $1.60 \pm 0.04$ in the other observations. We note that letting the photon index vary among the various flux levels resulted in consistent results. It is also clear that max-to-min variability due to the change in the power-law flux is $\sim 2.4$ on average, while the rest of the spectral variability, observed in the soft X-rays, is driven by the absorption changes. In fact, in Obs. 1, the neutral and the ionized absorbers are variable. The different flux states, during this observation, require an $N_{\rm H, N} = 34.4_{-6.8}^{+11.1}	\times 10^{22} ~\rm cm^{-2}$ for the neutral absorber, with a covering fraction that varies between 0.16 and 0.48. The mildly and the highly ionized absorbesr are both almost fully covering the source ($CF_{\rm W1} = 0.92_{-0.05}^{+0.04}$, $CF_{\rm W2} = 0.91_{-0.08}^{+0.05} $). The column density of the mildly ionized absorber is $N_{\rm H, W1} = 8.2_{-1.1}^{+1.7} \times 10^{21}~\rm cm^{-2}$ with an ionization level varying by more than an order of magnitude ($\log \xi_{\rm W1} = 0.14 - 1.28$). The highly ionized absorber with $ N_{\rm H,W2} = 3.81_{-0.43}^{+0.79}	\times 10^{22}~\rm cm^{-2}$ shows a more moderate variability in the ionization level $ \log \xi_{\rm W2} = 1.99-2.28 $) throughout this observation. 

However, the situation is different for Obs. 2+3. In these observations, which span $\sim 2.5$~days in total, the neutral absorber shows no variations, fully covering the source with $N_{\rm H, N} = 1.62_{-0.28}^{+0.53}	 \times 10^{22}~\rm cm^{-2}$. The best-fit suggests variable mildly-ionized absorption. The different flux states require an absorber with $N_{\rm H,W1} = 7.83_{-0.38}^{+0.97} \times 10^{22}~\rm cm^{-2}$, $\log \xi_{\rm W1} \sim 0.17-1.33$ and a covering fraction of $0.95_{-0.02}^{+0.01}$.  As for the \nustar\ observation, we found a column density of the neutral absorber of $N_{\rm H,N} = 5.95_{-0.91}^{+2.04}\times 10^{22}~\rm cm^{-2}$.
\begin{table}
\centering

\caption{Best-fit parameters of power law, thermal emission (${\tt apec_{1,2}}$),  and reflection. The normalizations are in units of $10^{-3}$, $10^{-5}$, or $10^{-4}~\rm photon~s^{-1}~cm^{-2}~keV^{-1}$, as indicated by subscripts. Temperatures are in keV.}

\begin{tabular}{lccc}
\hline \hline

Parameter	&		Low		&		Medium		&		High		\\ \hline
$kT_1$	&	$		$	&	$	0.16 \pm 0.01	$	&	$		$	\\
Norm$_{\rm apec1, -5}$	&	$		$	&	$	5.46_{-0.72}^{+1.29}	$	&	$		$	\\
$kT_2$	&	$		$	&	$	0.78_{-0.06}^{0.03}	$	&	$		$	\\
Norm$_{\rm apec2,-5}$	&	$		$	&	$	1.61_{-0.19}^{+0.13}	$	&	$		$	\\ \hline
	&		\multicolumn{3}{c}{Obs. 1}										\\
$\Gamma$	&	$		$	&	$	1.67_{-0.04}^{+0.02}	$	&	$		$	\\
Norm$_{\rm PL,-3}$	&	$	1.12_{-0.12}^{_0.14}	$	&	$	1.78_{-0.16}^{+0.17}	$	&	$	2.51_{-0.21}^{+0.25}	$	\\
Norm$_{\rm pexmon, -4}$	&	$		$	&	$	8.28_{-2.31}^{+0.69}	$	&	$		$	\\ \hline
	&		\multicolumn{3}{c}{Obs. 2+3}										\\
$\Gamma$	&	$		$	&	$	1.6\pm 0.04	$	&	$		$	\\
Norm$_{\rm PL, -3}$	&	$	1.12_{-0.04}^{+0.17}	$	&	$	1.85_{-0.03}^{+0.29}	$	&	$	2.78_{-0.09}^{+0.45}	$	\\
Norm$_{\rm pexmon, -4}$	&	$		$	&	$	6.61_{-1.19}^{+2.37}	$	&	$		$	\\ \hline
	&		\multicolumn{3}{c}{\nustar}										\\
Norm$_{\rm PL, -3}$	&	$	 -	$	&	$	1.38_{-0.05}^{+0.25}	$	&	$	2.56_{-0.10}^{+0.45}	$	\\
Norm$_{\rm pexmon, -4}$	&	$		$	&	$	6.61^{\rm tied}	$	&	$		$	\\ \hline\hline

\end{tabular}
\label{table:emissionpar}
\end{table}

\begin{table}
\centering

\caption{Intrinsic (unabsorbed) fluxes in units of $10^{-12}~\rm erg~s^{-1}~cm^{-2}$ for all the emission components in the $0.5-2/3-10$ keV ranges.}

\begin{tabular}{lccc}
\hline \hline
	&		Low		&		Medium		&		High		\\ \hline
Apec1	&	$		$	&	$	0.027/-	$	&	$		$	\\
Apec2	&	$		$	&	$	0.029-	$	&	$		$	\\ \hline
	&		\multicolumn{3}{c}{Obs. 1}										\\
Power law	&	$	2.59/3.82	$	&	$	4.07/6.01	$	&	$	5.79/8.56	$	\\
Pexmon	&	$		$	&	$	-/0.37	$	&	$		$	\\ \hline
	&		\multicolumn{3}{c}{Obs. 2+3}										\\
Power law	&	$	2.65/4.47	$	&	$	4.41/7.45	$	&	$	6.66/11.26	$	\\
Pexmon	&	$		$	&	$	-/0.41	$	&	$		$	\\ \hline
	&		\multicolumn{3}{c}{\nustar}										\\
Power law	&	$	-	$	&	$	-/5.59	$	&	$	-/10.03	$	\\ \hline\hline

\end{tabular}
\label{table:fluxes}
\end{table}

The thermal soft components ({\tt apec}) are conservatively assumed to be constant for all observations (see Section \ref{sec:softemission} for more details about possible degeneracies and caveats in modeling the soft emission). The best-fit temperatures are 0.16~keV and 0.78~keV. The $0.5-2$~keV and $3-10$~keV fluxes of each emission component are listed in Table~\ref{table:fluxes}. The sum of the soft components ({\tt apec1,2}) results in a flux of $\sim 5.6\times 10^{-14}~\rm erg~s^{-1}~cm^{-2}$ that is $\sim 1-2\%$ of the intrinsic power law flux in the $0.5-2$~keV range. We also note that the reflected flux is constant during all observations. The intrinsic unabsorbed power-law luminosity in the $3-10$~keV range varies between $9.1 \times 10^{39}~\rm erg~s^{-1}$ and $2.4 \times 10^{40}~\rm erg~s^{-1}$.


\section{Discussion}
\label{sec:discussion}

We have analyzed multi-epoch \xmm\ and \nustar\ flux-resolved spectra of the low-luminosity highly-variable Seyfert galaxy NGC 4395. Our modeling suggests that the nuclear emission is obscured by three layers of absorption: neutral, mildly ionized, and highly ionized. The extent of intrinsic variability (a factor of $\sim 2.5$) is revealed by the hard X-rays as probed by \nustar\ and cannot by itself account for all the flux variation (a factor of more than 10) that is observed in the soft X-rays (below 2 keV) during Obs. 1. To quantitatively estimate the expected variability within the context of our spectral modeling, we considered the best-fit model ({\tt M3}) to simulate two sets of light curves. First, we removed all absorption components from {\tt M3}, and created 2000 \xmm\ light curves, with an exposure time of 1~ks each, assuming the best-fit parameters corresponding to Obs. 1. We considered the change in PL normalization as being the only source of variability. We assumed a log-normal distribution of the PL normalization that is consistent with the distribution of the 2-10 keV count rates. The black dash-dotted line in Fig. \ref{fig:Fvar} corresponds to the estimated $F_{\rm var}$ for this scenario. $F_{\rm var}$ is almost constant ($\langle F_{\rm var} \sim 0.3 \rangle$) over the full the $0.4-10$~keV range, except for a clear dip in the $6-7$~keV range, characterized by the constant Fe line in the {\tt pexmon} model. We repeated the same experiment by considering the best-fit {\tt M3} parameters from Obs. 2+3 and considering a constant neutral absorption with $N_{\rm H} = 1.6\times 10^{22}~\rm cm^{-2}$. The $F_{\rm var}$ estimated for both \xmm\ and \nustar\ is shown in Fig.~\ref{fig:Fvar} (dashed black lines and green dotted lines, respectively). The observed $0.4-1.5$~keV band is dominated by the constant components (${\tt apec_{1,2}}$), which reduce the observed variability in this range. Above 1.5~keV, $F_{\rm var}$ follows a similar behavior to the previous set of simulations (with no absorption). Interestingly, a small decrease in the $F_{\rm var}$ is also seen in the $\sim 15-30$~keV corresponding to the Compton-hump of the {\tt pexmon} component. The simulations are in agreement with the measured $F_{\rm var}$ below 1~keV in Obs. 2+3, and above 4~keV for all observations. However, an excess in $F_{\rm var}$ can be seen in the $0.4-4$~keV and $1-4$~keV ranges in Obs. 1 and Obs. 2+3, respectively. We attribute this additional variability in Obs. 1 to independent (random) changes in the covering fraction of the neutral absorber and the ionization level of the ionized absorbers. This is consistent with the time-resolved spectral analysis (NR11) and PCA analysis \citep{Parker15} of Obs. 1. 

As NR11 mentioned, the eclipse-like events seen in Obs. 1 could be either due to a single, inhomogeneous cloud or a system of different small clouds. The absorption variability might act on longer timescales compared to the intrinsic one, which will lead to a shift in the characteristic timescale towards smaller values for higher energy ranges (as seen in Fig. \ref{fig:PSD}). As for Obs.2+3, the neutral absorber shows no variations and fully covers the source explaining the low variability level that is observed below 1~keV. We note that additional variability seen in the 1-4~keV range is probably due to changes in the ionization level of mildly ionized absorber. The fact that the characteristic timescales are consistent at all energies in the PSDs of Obs. 2+3 might indicate that either a) the intrinsic flux change and the ionization level changes are acting on similar timescales, or b) the effects of the change in ionization level are small compared to the ones due to the intrinsic flux change and could not be identified by the current data quality.

We stress that the flux-resolved analysis, presented in this work, likely probes the full range of variability of the different components. This is complementary to the time-resolved analysis probing the succession of different states. This is more relevant for Obs. 1 which shows a more complex behavior than Obs. 2+3, where the neutral absorber shows no variability. Our results give the characteristic absorption/emission properties required by each flux state. In our approach, the flux levels are defined from the $2-10$ keV band showing moderate variability, since it is less affected by absorption compared to lower energies. Hence, any correlation between flux state and the obscuration level is not obvious a priori. For instance, the covering fractions of the neutral low- and medium-flux states in Obs. 1 are consistent. However, it could be possible that at high flux levels the gas becomes more ionized, hence the impact of neutral absorption diminishes. Testing this hypothesis requires an accurate identification of all ionization phases, tracking also the evolution of $N_{\rm H}$ and the covering fraction of each of them. This will be possible with the next generation of X-ray observatories. Any study of the absorber's structure and its evolution requires a time-resolved approach similar to NR11. We finally note that our results are qualitatively in agreement with the findings of NR11. The variability due to absorption is associated with the higher column density absorber (of the order $10^{23}~\rm cm^{-2}$). The lower column density absorber (of the order $10^{22}~\rm cm^{-2}$) is less variable, with the main difference compared to NR11 being that the current analysis finds this absorber to almost fully cover the source.

Interestingly, \cite{McHardy16} found that the X-rays lead the UVW1 and the $g$-band light curves by 473~s and 788~s, respectively, during Obs. 2+3. This indicates that the UV/optical reprocessing region responding to the X-ray variability on these timescales is located at a distance that is closer to the BH than the BLR. This is consistent with thermal reprocessing by a standard accretion disk \citep[see e.g.,][]{Cackett07, Kammoun19}. However, the current data quality does not allow us to identify any ionized reflection from the accretion disk in the X-ray spectra. 

\section{The soft X-ray emission}
\label{sec:softemission}
We acknowledge that our modeling of the soft X-rays is limited by the low spectral resolution at low energies. We model the soft X-ray spectra by including two {\tt apec} components. We associate the {$\tt apec_1$} component to the extended soft emission regions seen in the {\it Chandra} and the {\it HST} [\ion{O}{3}] images \citep[see e.g.,][]{Gomez17}. However, the nature of the {$\tt apec_2$} component is uncertain. This component could be the emission counterpart of the mildly-ionized absorber or simply accounts for the gradient of temperature in the diffuse gas. But its presence could be compensating for some inadequacy of the employed absorption (and emission) models (e.g., the residuals at $\sim 0.75$~keV in the high flux spectrum of Obs. 1). We also tested the possibility that the soft X-ray emission is due to a smooth soft component (modeled with a blackbody), which could be associated with the intrinsic disk emission given the low BH mass, in addition to a thermal diffuse emission (modeled with an {\tt apec} component). This results in a statistically worse fit with $\rm \chi^2/dof = 1594/1530$ ($\rm \Delta \chi^2  = +26$ for the same dof). In fact, as the intrinsic X-ray emission is obscured, it would be natural that parts of the accretion disk (mainly the innermost region contributing most to the soft X-ray emission) are also obscured. However, it is impossible to identify which parts of the disk are obscured and their covering fractions, knowing that the obscuring material has a complex and inhomogeneous structure. Overcoming the uncertainties in modeling the soft X-ray spectrum of this source requires higher spectral resolution, as will be provided by the next generation of X-ray missions. These missions would help accurately identify any thermal emission and/or absorption structures (see Section \ref{sec:futuremissions}).

\begin{figure}
\centering

\includegraphics[width = 1.\linewidth]{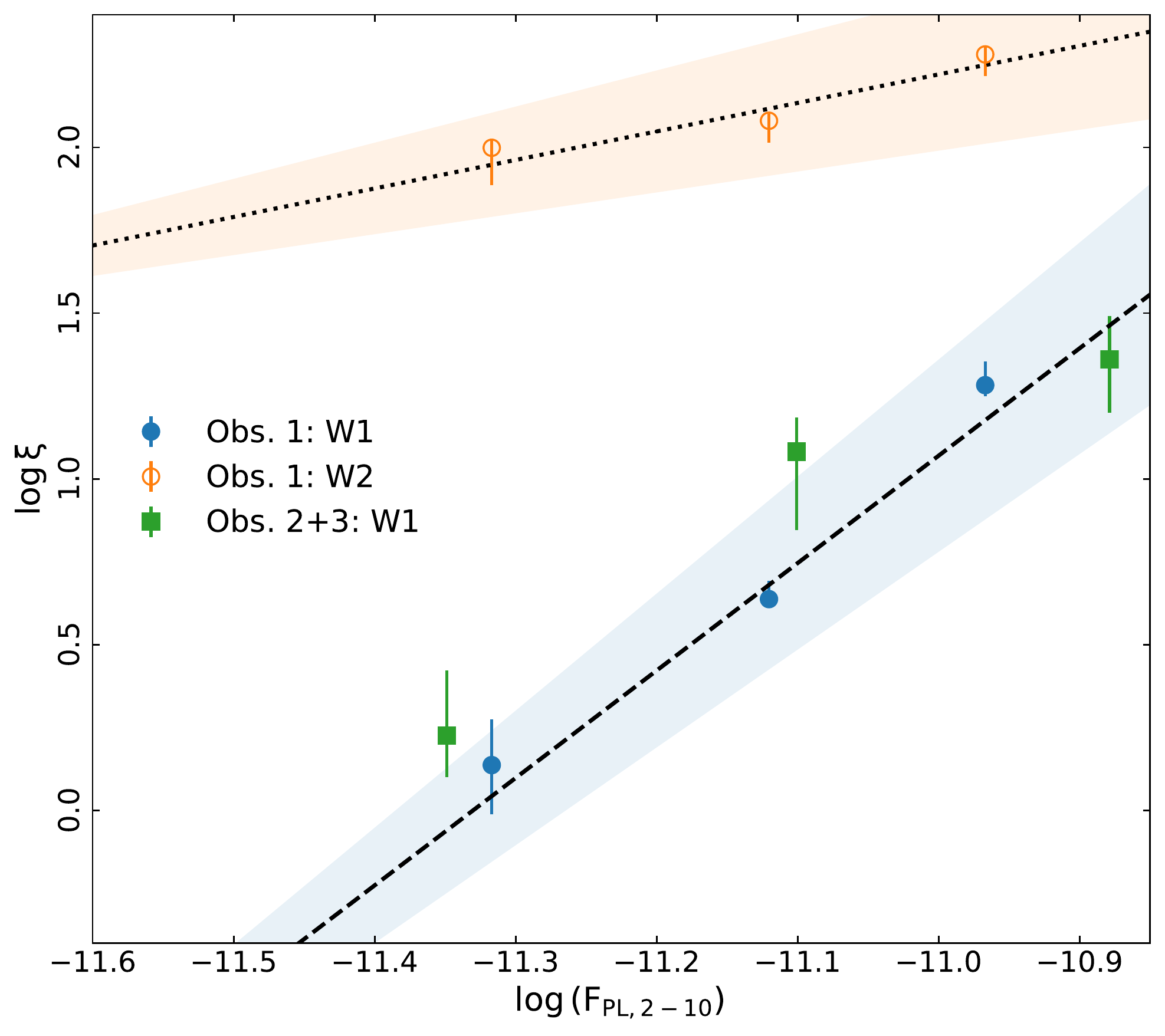}

\caption{Best-fit ionization parameter (in units of $\rm erg~cm~s^{-1}$) as function of the power law flux in the $3-10$~keV band (in units of $\rm erg~s^{-1}~cm^{-2}$), for all observations. The solid and dashed lines are the best-fit linear relation for Obs. 1 and Obs. 2+3, respectively. The error bars and the shaded regions correspond to the 1$\sigma$ confidence level on the fitted ionization level and the linear relationship, respectively.}
\label{fig:logxi-norm}
\end{figure}

\subsection{The variable ionized absorption}
\label{sec:WA}

Our results show that the ionized absorbers (both mildly and highly ionized) vary as a function of flux. We assume in our analysis that the variability is just in the ionization level of these components. We test the validity of this hypothesis by tying different pairs in the column density, ionization level, and covering fraction space, letting the third parameter free to vary. We found that letting the column density or the covering fraction free to vary results in statistically unacceptable fits with $\chi^2 = 1599$ and $1651$ for 1530 dof, respectively, compared to $\chi^2 = 1568$ for a free ionization level with the same dof. Figure~\ref{fig:logxi-norm} shows the ionization level of the mildly (Obs. 1 and 2+3) and highly (Obs. 1 only) ionized absorbers as a function of the intrinsic power-law flux. This figure shows a clear positive correlation between the two quantities, for the two absorbers. This may indicate that the ionized absorbers are responding to the flux changes on timescales that are comparable to the intrinsic variability timescale of the power law emission. We fit the $\log \xi$ versus $\log F$ points for Obs. 1 and Obs. 2+3 together for the mildly ionized absorber (dashed line) and the highly ionized absorber (dotted line) assuming a linear correlation. The slopes of the correlations are $3.2 \pm 0.3$ and $0.9 \pm 0.2 $ for the mildly and highly ionized absorbers, respectively. The slope of the highly ionized absorber is consistent with unity, as expected from the definition of the ionization parameter for a constant density and location. However, for the mildly ionized absorber, the slope of the correlation is larger than unity, which may indicate some change in the location and/or the density of the absorbing material. However, neither the size (hence the density derived from $N_{\rm H}$) nor the location of the gas could be well determined using the current data quality. We note that a similar relation between the ionization level and the intrinsic flux has been seen in other objects, for instance NGC 4151 \citep[][]{Schurch02, Zoghbi19}.

\subsection{The BLR size}
\label{sec:BLR}

The best-fit model suggests that the neutral reflection is constant among all observations. Thus, we can use this component as a tracer of the innermost extent of the cold obscuring material, assuming that it is responsible for the observed reprocessed emission. We added an {\tt rdblur} relativistic blurring function to modify the {\tt pexmon} component. This model assumes a Schwarzschild black hole, and a power-law emissivity profile ($\epsilon \propto r^{-q}$). We fixed the emissivity index at $q =3$ and the outermost radius of the material at $10^6~\rm r_g$ (where $\rm r_g = GM_{BH}/c^2$ is the gravitational radius), considering a fixed inclination of 45\degr. We obtained a lower limit on the innermost radius $R_{\rm in} \gtrsim 4600~\rm r_g$. This corresponds to $6.9\times 10^{12}~\rm cm$ or $2.5 \times 10^{14}~\rm cm$ for $M_{\rm BH} = 10^4~\rm M_\odot$ or $M_{\rm BH} =3.5 \times 10^5~\rm M_\odot$, respectively. The fit is driven mainly by the narrow Fe line, disfavoring a broader line profile hence smaller value of $R_{\rm in}$. Using the ionization parameter definition ($\xi = L/nr^2$), we can get a rough estimate of the distance of the neutral material. Assuming an ionizing luminosity of $\sim 10^{41}~\rm erg~s^{-1}$ and a density of $10^{9-11}~\rm cm^{-3}$, we obtain $r \geq 10^{15-16}~\rm cm$. This estimate is broadly consistent with the value obtained by blurring the reflection component, and with the typical size of the BLR ($R_{\rm BLR}$). In fact, \cite{Peterson05} and \cite{Woo19} determined a time lag of $\sim 1~\rm hr$ and $\sim 1.4~\rm hr$  for the \ion{C}{4} $\lambda1549$ and H$\alpha$ emission lines, respectively, which gives $R_{\rm BLR} \sim 10^{14}~\rm cm$. This corresponds to $6.7\times 10^4 ~ (1.9\times 10^3) ~\rm r_g$ for $M_{\rm BH} = 10^4~(3.6\times 10^5)~ \rm M_\odot$.
\begin{table}

\caption{The distance of the BLR and the sizes of the obscuring clouds assuming different masses and line velocity dispersions.}
\begin{tabular}{lrrrr}
\hline \hline
$M_{\rm BH}	$&		&		&	$ 10^4~\rm M_\odot$	&	$3.6\times 10^5~\rm M_\odot$	\\ \hline
$R_{\rm BLR}$	&		&	$10^{14}~\rm cm$	&	$6.7 \times 10^4~\rm r_g$	&	$1.9 \times 10^3~\rm r_g$	\\ \hline
\multicolumn{5}{l}{Obs. 1 ($\delta t = 10~\rm ks$)}									\\ \\[-0.2cm]
$D_{\rm C}$	&	$\sigma = 500~\rm km~s^{-1}$	&	$5 \times 10^{11}~\rm cm$	&	$333 ~\rm r_g$	&	$9.26 ~\rm r_g$	\\
	&	$\sigma = 1500~\rm km~s^{-1}$	&	$1.5 \times 10^{12}~\rm cm$	&	$1000 ~\rm r_g$	&	$28 ~\rm r_g$	\\ \hline
\multicolumn{5}{l}{Obs. 2+3 ($\delta t > 220~\rm ks$)}									\\ \\[-0.2cm]
$D_{\rm C}$	&	$\sigma = 500~\rm km~s^{-1}$	&	$1.1 \times 10^{13}~\rm cm$	&	$7.3 \times 10^3~\rm r_g$	&	$200 ~\rm r_g$	\\
	&	$\sigma = 1500~\rm km~s^{-1}$	&	$3.3 \times 10^{13}~\rm cm$	&	$2.2 \times 10^4 ~\rm r_g$	&	$611 ~\rm r_g$	\\ \hline \hline
\end{tabular}
\label{table:distances}
\end{table}

The difference in the behavior of the neutral absorber between Obs.1 (short timescale variability) and Obs. 2+3 (constant over longer timescales) might indicate an inhomogeneous and layered BLR. A simple calculation can be done in order to have an estimate of the size of the obscuring neutral clouds. The variability timescale associated with the change in obscuration can be approximated as $\delta t \simeq D_{\rm C} / v$, where $D_{\rm C}$ is the characteristic size of the obscuring material, and $v$ is the transverse velocity of the medium. Obs. 1 shows clear variations in the neutral obscuring material. NR11 argue that, during this observation, the source exhibited eclipse-like events on timescales of $\sim 10~\rm ks$. This is also confirmed by our analysis that shows variations in the covering fraction between the low-/medium-flux (0.48/0.38) states and the high-flux state (0.16). However, the neutral absorber remains constant and fully covering the source during Obs. 2+3. This can give us only a lower limit on the size of the obscuring material during these observations. The exact location and velocity of the obscuring material is unknown. \cite{Peterson05} reported a velocity dispersion $\sigma \sim 1500~\rm km~s^{-1}$ for \ion{C}{4}, while \cite{Woo19} reported a value of $\sigma \sim 426~\rm km~s^{-1}$ for the H$\alpha$ line. Given this, and the uncertainty on the mass measurement, we estimate the cloud size assuming $v = 500~\rm km~s^{-1}$ and $1500~\rm km~s^{-1}$, and the two mass measurements reported in the literature, as shown in Table~\ref{table:distances}. The values are reported in cm and in $\rm r_g$. We assume a variability timescale of 10~ks for Obs. 1, while for Obs. 2+3 we can only estimate a lower limit on the size of the cloud, assuming $\delta t > 220~\rm ks$. It is more likely that the absorber in Obs.1 is smaller, faster and closer to the source compared to the absorber in Obs. 2+3. In these observations, the obscuration is most likely due to a slower and bigger single cloud, located at a larger distance. Assuming that the cloud density is $n_{\rm H} \sim N_{\rm H}/ D_{\rm C}$ we get densities of $\sim 2\times 10^{11}~\rm cm^{-3}$ for Obs. 1 (assuming $\sigma = 1500~\rm km~s^{-1}$) and $\sim 10^9~\rm cm^{-3}$ for Obs. 2+3 (assuming $\sigma = 500~\rm km~s^{-1}$), given the best-fit $N_{\rm H,N}$ values listed in Table~\ref{table:absorptionpar}. 

\begin{figure*}
\centering

\includegraphics[width = 0.99\linewidth]{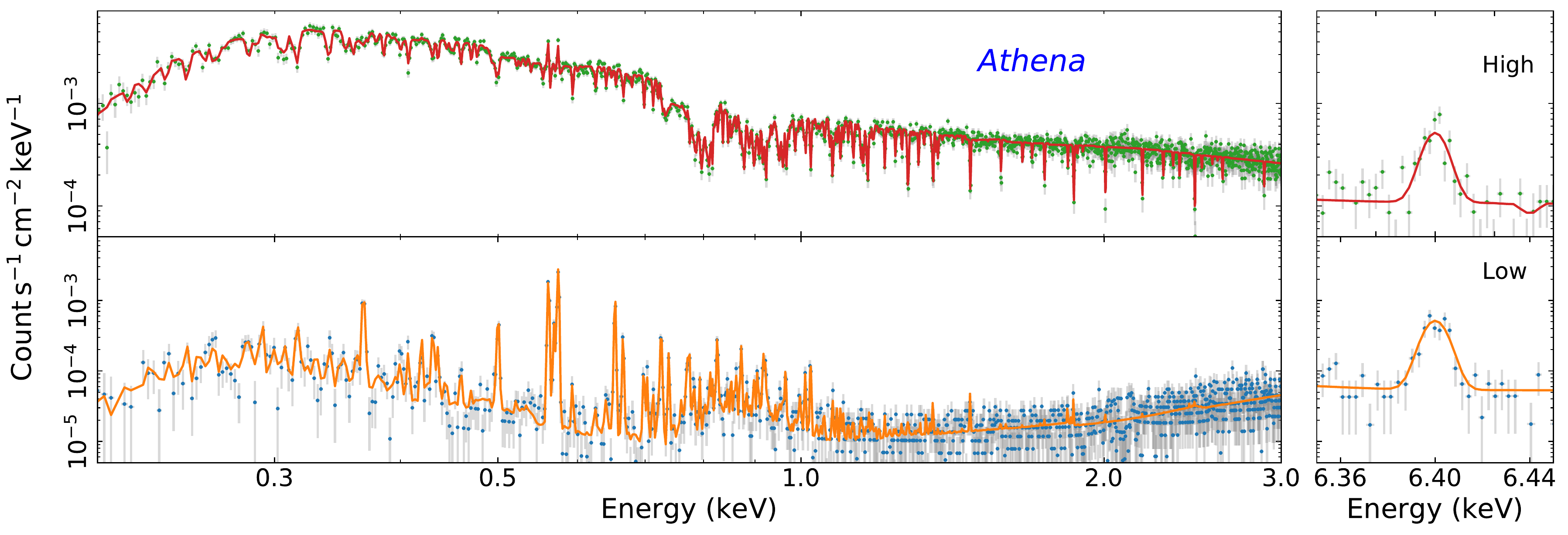}
\includegraphics[width = 0.99\linewidth]{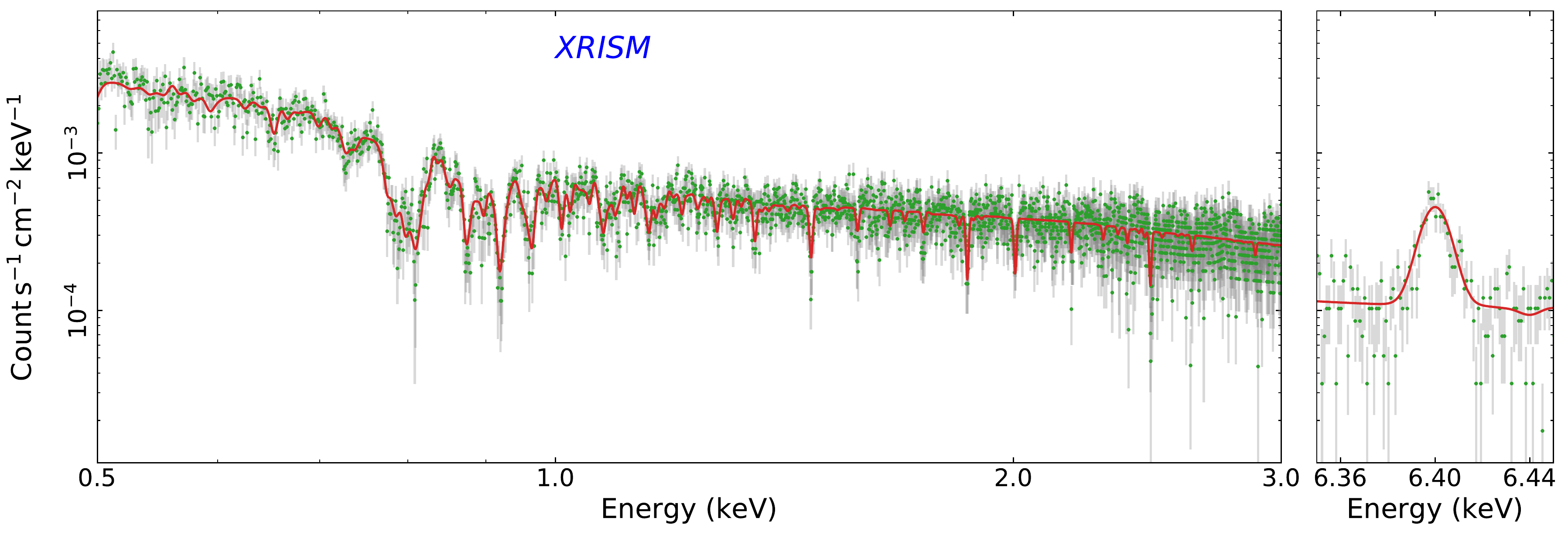}

\caption{Top: Simulated 10 ks {\it Athena}/X-IFU spectra of NGC 4395 assuming the high-flux Obs. 1 (upper row) and low-flux Obs. 2 (bottom row) best-fit models (solid lines). The spectra are binned using the `optimal' binning scheme \citep{Kaastra16}. Bottom: Simulated 200 ks {\it XRISM}/Resolve spectra of NGC 4395 assuming the high-flux Obs. 1 best-fit model (red solid line). The RMF files are linearly compressed reducing the number of channels by a factor of 2. No grouping is applied to the spectra. The low-flux Obs. 2+3 state  is dominated by the background, so we did not include it in the simulations. The left and right panels show the spectra below 3~keV and in the Fe K$\alpha$ band, respectively.}
\label{fig:xrismathena}
\end{figure*}

It is possible that the neutral absorption in principle could be related some outflowing material from the disk which could also extend to the BLR, as seen in NGC 5548 for example \citep{Kaastra14}. In that case, it might be possible that the ionization state of the absorption in this source shows a radial dependence. Testing this would require high-resolution UV and X-ray spectra that would allow to measure the ionization state of this material and energy shifts due to its motion. This would be possible with the next generation of X-ray observatories. The layered geometry, proposed in this work, is consistent with the general notion of a clumpy BLR and torus, as seen in several sources \citep{Risaliti05, Bianchi12, Miniutti14}. The rapid changes seen in the neutral absorption during Obs. 1 are consistent with originating from the BLR. The lack of variability in the neutral absorption during Obs. 2+3 allows us to infer lower limits only. Hence, it is possible that the neutral material obscuring the source in these observations is located in the outer extent of the BLR or even in the torus \citep[see e.g.,][]{Miniutti14}.

\subsection{Future missions}
\label{sec:futuremissions}
\begin{figure}
\centering

\includegraphics[width = 0.99\linewidth]{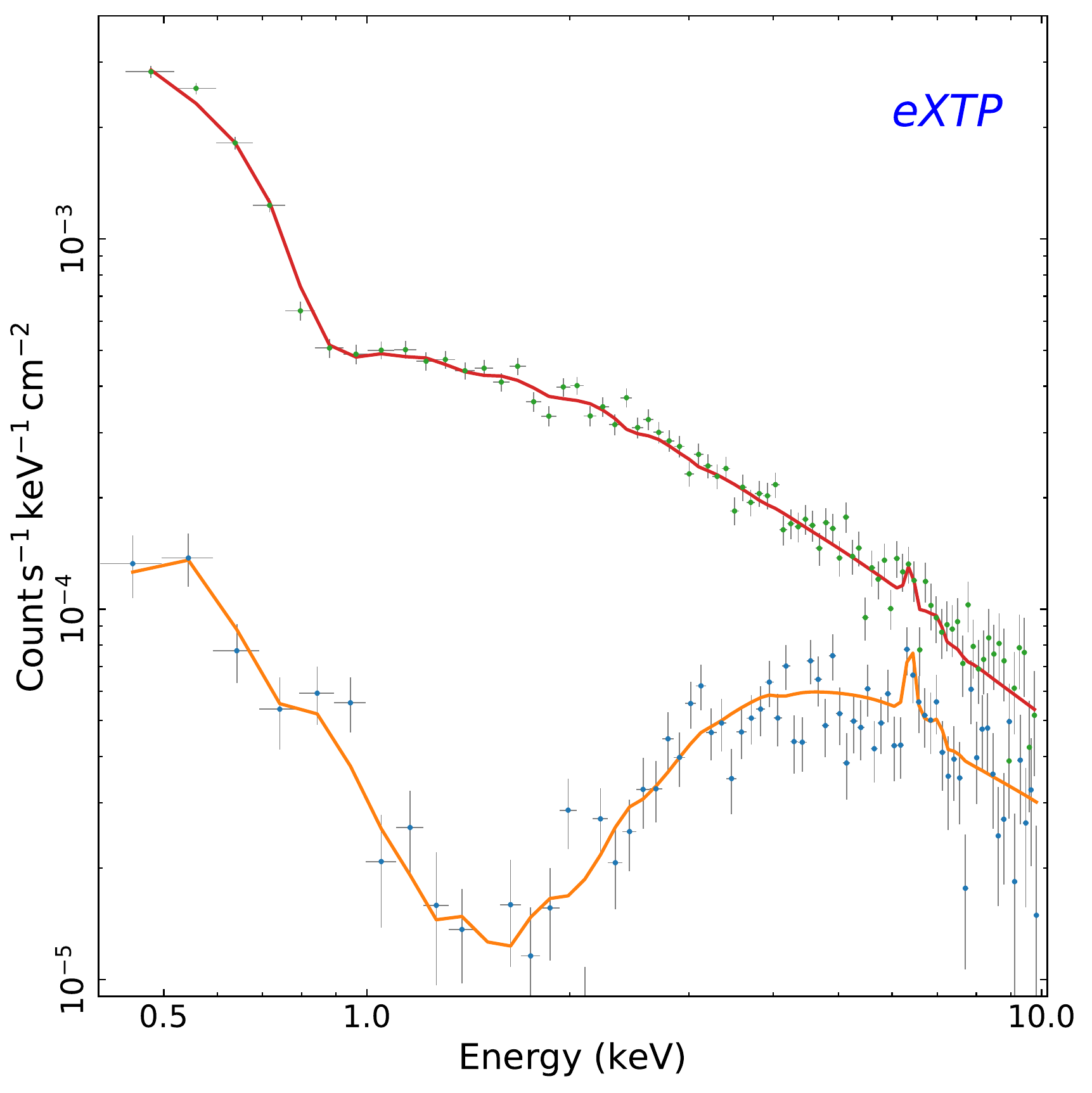}

\caption{Same as Fig.~\ref{fig:xrismathena} but for the {\it eXTP}/SFA, with an exposure time of 1~ks for each spectrum. The spectra are binned using the `optimal' binning scheme.}
\label{fig:eXTP}
\end{figure}

Given the potential layering in the BLR inferred from our current analysis, future X-ray missions would help determine the low/moderate ionization state of the absorbers which will allow us to better locate the different structures along the line of sight, hence determining their orbital velocities and sizes. The high spectral resolution and sensitivity of future X-ray missions will allow us to identify many absorption and emission spectral features. This would be crucial to understanding the nature of the soft X-ray features: whether they are caused by a complex diffuse thermal emission or a smooth absorbed blackbody. Furthermore, future missions would enable a better understanding  of the nature and the dynamics of the obscuring material and some better insights on outflows and AGN feedback. It would also be possible to identify the different ionization phases of the BLR gas and study the evolution of their absorbing columns and covering fractions. The bottom panel of Fig.~\ref{fig:xrismathena} shows simulated 200-ks {\it XRISM}/Resolve\footnote{\url{https://heasarc.gsfc.nasa.gov/docs/xrism/proposals/}} \citep{Xrism} spectrum based on the high-flux Obs. 1 best-fit model. The low-flux level of this source, based on our fits, would be comparable to the background level of the instrument. We stress that the simulated {\it XRISM} high-flux spectrum does not correspond to a 200-ks observation, but instead it is observing the source in its highest flux state for that exposure, which will require a long monitoring. Instead, {\it XRISM} will allow us to study the Fe line profile even when the source is in its lowest flux state. This would allow us to determine the geometry and the location of the reprocessing material. In addition, it would help identify any disk reflection component and its possible eclipse.

Eclipsing events could be also studied by the `enhanced X-ray Timing and Polarimetry' mission \citep[\textit{eXTP},][]{eXTP19}. Thanks to the relatively large effective area of the Spectroscopic Focusing Array (SFA) we will be able to obtain spectra on timescales as short as 1~ks as shown in Fig.~\ref{fig:eXTP}, though with low energy resolution, that would be appropriate for the fast variability and the characteristic timescales for this source (see Section~\ref{sec:psd}). For an eclipse timescale in the order of $\sim 10$~ks we will be able to probe the full evolution of the covering fraction. We also note that a crossing time $\delta t = 1$~ks would correspond to a distance of $\sim 2\times 10^4~\rm r_g$ or $560~\rm r_g$ for $M_{\rm BH} = 10^4~\rm M_\odot$ or $3.6\times 10^5~\rm M_\odot$, respectively. 

{\it Athena}/X-IFU \citep{XIFU} would allow us to combine the time and energy resolution. The top panel of Fig.~\ref{fig:xrismathena} shows simulated 10-ks {\it Athena}/X-IFU spectra\footnote{\url{http://x-ifu.irap.omp.eu/resources-for-users-and-x-ifu-consortium-members/}} based on the high-flux Obs. 1 and low-flux Obs. 2 best fits. X-IFU will allow us to study the variability in absorption on short timescales with a high accuracy \citep[see][for more details about the ability of X-IFU to study absorption features in AGN]{Barret19}. In addition, this will allow us to reveal any possible variability in the reflection spectrum, or broadening of the Fe K$\alpha$ feature which would help identify the location and the nature of the reprocessing material. High-resolution spectra would help in tracking any eclipsing events in NGC 4395, allowing us to probe the innermost region close to the BH \citep[see][for more details]{Kammoun18}.

\section{Conclusion}
\label{sec:conclusion}

We have presented  a detailed flux-resolved analysis of the X-ray spectra of NGC 4395 using multi-epoch non-simultaneous \xmm\ and \nustar\ observations. Our results suggest that the source is affected by a complex structure of absorbing material. It consists of three layers with different ionization levels: neutral, mildly ionized, and highly ionized. The neutral material shows variations in its covering fraction during Obs. 1, which, in addition to the intrinsic variability, explains the high observed variability (a factor of $\sim 6$) in the soft X-rays. However, this layer remains constant during Obs. 2+3 where the variability is mainly intrinsic. The intrinsic variability could be also detected in the hard X-rays with \nustar. The ionization level of both midly and highly ionized absorbers increase with the intrinsic flux, which indicates a response of the flux changes on timescales comparable to the intrinsic timescale. Our spectral modeling is also supported by the dependence of the PSD and $F_{\rm var}$ on energy. Future missions would allow us to study in detail the absorption/emission features in addition to the evolution of any absorption changes on their typical timescales.

\begin{acknowledgements}
This work made use of data from the \nustar\ mission, a project led by the California Institute of Technology, managed by the Jet Propulsion Laboratory, and funded by NASA,
\xmm, an ESA science mission with instruments and contributions directly funded by ESA Member States and NASA. This research has made use of the \nustar\ Data Analysis Software (NUSTARDAS) jointly developed by the ASI Science Data Centre (ASDC, Italy) and the California Institute of Technology (USA).The figures were generated using matplotlib \citep{Hunter07}, a {\tt PYTHON} library for publication of quality graphics. The MCMC results were presented using the GetDist {\tt PYTHON} package.
\end{acknowledgements}

\software{EMCEE \citep{EMCEE13}, HEASoft \citep{HEASoft}, Matplotlib \citep{Hunter07}, NUSTARDAS (v1.8.0, \url{https://heasarc.gsfc.nasa.gov/docs/nustar/analysis/}, SAS  \citep[v17.0.0][]{Gabriel04}, XSPEC \citep{Arnaud96}, XSPEC\_EMCEE (\url{https://github.com/jeremysanders/xspec_emcee}).}

\facilities{\nustar, \xmm.}
	\bibliographystyle{aasjournal}
	\bibliography{ek-NGC4395} 

\begin{thebibliography}{}
\expandafter\ifx\csname natexlab\endcsname\relax\def\natexlab#1{#1}\fi
\providecommand{\url}[1]{\href{#1}{#1}}
\providecommand{\dodoi}[1]{doi:~\href{http://doi.org/#1}{\nolinkurl{#1}}}
\providecommand{\doeprint}[1]{\href{http://ascl.net/#1}{\nolinkurl{http://ascl.net/#1}}}
\providecommand{\doarXiv}[1]{\href{https://arxiv.org/abs/#1}{\nolinkurl{https://arxiv.org/abs/#1}}}

\bibitem[{{Arnaud}(1996)}]{Arnaud96}
{Arnaud}, K.~A. 1996, in Astronomical Society of the Pacific Conference Series,
  Vol. 101, Astronomical Data Analysis Software and Systems V, ed. G.~H.
  {Jacoby} \& J.~{Barnes}, 17

\bibitem[{{Barret} \& {Cappi}(2019)}]{Barret19}
{Barret}, D., \& {Cappi}, M. 2019, \aa, arXiv:1906.02734.
\newblock \doarXiv{1906.02734}

\bibitem[{{Barret} \& {Vaughan}(2012)}]{Barret12}
{Barret}, D., \& {Vaughan}, S. 2012, \apj, 746, 131,
  \dodoi{10.1088/0004-637X/746/2/131}

\bibitem[{{Barret} {et~al.}(2018){Barret}, {Lam Trong}, {den Herder}, {Piro},
  {Cappi}, {Houvelin}, {Kelley}, {Mas-Hesse}, {Mitsuda}, \& {Paltani}}]{XIFU}
{Barret}, D., {Lam Trong}, T., {den Herder}, J.-W., {et~al.} 2018, in Society
  of Photo-Optical Instrumentation Engineers (SPIE) Conference Series, Vol.
  10699, Space Telescopes and Instrumentation 2018: Ultraviolet to Gamma Ray,
  106991G

\bibitem[{{Belloni} {et~al.}(2002){Belloni}, {Psaltis}, \& {van der
  Klis}}]{Belloni02}
{Belloni}, T., {Psaltis}, D., \& {van der Klis}, M. 2002, \apj, 572, 392,
  \dodoi{10.1086/340290}

\bibitem[{{Belloni} {et~al.}(1997){Belloni}, {van der Klis}, {Lewin}, {van
  Paradijs}, {Dotani}, {Mitsuda}, \& {Miyamoto}}]{Belloni97}
{Belloni}, T., {van der Klis}, M., {Lewin}, W.~H.~G., {et~al.} 1997, \aap, 322,
  857

\bibitem[{{Bianchi} {et~al.}(2012){Bianchi}, {Maiolino}, \&
  {Risaliti}}]{Bianchi12}
{Bianchi}, S., {Maiolino}, R., \& {Risaliti}, G. 2012, Advances in Astronomy,
  2012, 782030, \dodoi{10.1155/2012/782030}

\bibitem[{{Blackburn}(1995)}]{Blackburn95}
{Blackburn}, J.~K. 1995, in Astronomical Society of the Pacific Conference
  Series, Vol.~77, Astronomical Data Analysis Software and Systems IV, ed.
  R.~A. {Shaw}, H.~E. {Payne}, \& J.~J.~E. {Hayes}, 367

\bibitem[{{Cackett} {et~al.}(2007){Cackett}, {Horne}, \& {Winkler}}]{Cackett07}
{Cackett}, E.~M., {Horne}, K., \& {Winkler}, H. 2007, \mnras, 380, 669,
  \dodoi{10.1111/j.1365-2966.2007.12098.x}

\bibitem[{{De Marco} {et~al.}(2013){De Marco}, {Ponti}, {Cappi}, {Dadina},
  {Uttley}, {Cackett}, {Fabian}, \& {Miniutti}}]{Demarco13}
{De Marco}, B., {Ponti}, G., {Cappi}, M., {et~al.} 2013, \mnras, 431, 2441,
  \dodoi{10.1093/mnras/stt339}

\bibitem[{{Dewangan} {et~al.}(2008){Dewangan}, {Mathur}, {Griffiths}, \&
  {Rao}}]{Dewangan08}
{Dewangan}, G.~C., {Mathur}, S., {Griffiths}, R.~E., \& {Rao}, A.~R. 2008,
  \apj, 689, 762, \dodoi{10.1086/591728}

\bibitem[{{Filippenko} {et~al.}(1993){Filippenko}, {Ho}, \&
  {Sargent}}]{Filippenko93}
{Filippenko}, A.~V., {Ho}, L.~C., \& {Sargent}, W. L.~W. 1993, \apjl, 410, L75,
  \dodoi{10.1086/186883}

\bibitem[{{Filippenko} \& {Sargent}(1989)}]{Filippenko89}
{Filippenko}, A.~V., \& {Sargent}, W. L.~W. 1989, \apj, 342, L11,
  \dodoi{10.1086/185472}

\bibitem[{{Foreman-Mackey} {et~al.}(2013){Foreman-Mackey}, {Hogg}, {Lang}, \&
  {Goodman}}]{EMCEE13}
{Foreman-Mackey}, D., {Hogg}, D.~W., {Lang}, D., \& {Goodman}, J. 2013, \pasp,
  125, 306, \dodoi{10.1086/670067}

\bibitem[{{Gabriel} {et~al.}(2004){Gabriel}, {Denby}, {Fyfe}, {Hoar}, {Ibarra},
  {Ojero}, {Osborne}, {Saxton}, {Lammers}, \& {Vacanti}}]{Gabriel04}
{Gabriel}, C., {Denby}, M., {Fyfe}, D.~J., {et~al.} 2004, in Astronomical
  Society of the Pacific Conference Series, Vol. 314, Astronomical Data
  Analysis Software and Systems (ADASS) XIII, ed. F.~{Ochsenbein}, M.~G.
  {Allen}, \& D.~{Egret}, 759

\bibitem[{{Garc{\'{\i}}a} {et~al.}(2013){Garc{\'{\i}}a}, {Dauser}, {Reynolds},
  {Kallman}, {McClintock}, {Wilms}, \& {Eikmann}}]{Garcia13}
{Garc{\'{\i}}a}, J., {Dauser}, T., {Reynolds}, C.~S., {et~al.} 2013, \apj, 768,
  146, \dodoi{10.1088/0004-637X/768/2/146}

\bibitem[{{Garc{\'{\i}}a} \& {Kallman}(2010)}]{Garcia10}
{Garc{\'{\i}}a}, J., \& {Kallman}, T.~R. 2010, \apj, 718, 695,
  \dodoi{10.1088/0004-637X/718/2/695}

\bibitem[{{G{\'o}mez-Guijarro} {et~al.}(2017){G{\'o}mez-Guijarro},
  {Gonz{\'a}lez-Mart{\'\i}n}, {Ramos Almeida}, {Rodr{\'\i}guez-Espinosa}, \&
  {Gallego}}]{Gomez17}
{G{\'o}mez-Guijarro}, C., {Gonz{\'a}lez-Mart{\'\i}n}, O., {Ramos Almeida}, C.,
  {Rodr{\'\i}guez-Espinosa}, J.~M., \& {Gallego}, J. 2017, \mnras, 469, 2720,
  \dodoi{10.1093/mnras/stx1037}

\bibitem[{{Gonz{\'a}lez-Mart{\'\i}n}(2018)}]{GonzalezMartin18}
{Gonz{\'a}lez-Mart{\'\i}n}, O. 2018, \apj, 858, 2,
  \dodoi{10.3847/1538-4357/aab7ec}

\bibitem[{{Goodman} \& {Weare}(2010)}]{Goodman10}
{Goodman}, J., \& {Weare}, J. 2010, Comm. App. Math. Comp. Sci., 5, 65,
  \dodoi{10.2140/camcos.2010.5.65}

\bibitem[{{Harrison} {et~al.}(2013){Harrison}, {Craig}, {Christensen},
  {Hailey}, {Zhang}, {Boggs}, {Stern}, {Cook}, {Forster}, {Giommi},
  {Grefenstette}, {Kim}, {Kitaguchi}, {Koglin}, {Madsen}, {Mao}, {Miyasaka},
  {Mori}, {Perri}, {Pivovaroff}, {Puccetti}, {Rana}, {Westergaard}, {Willis},
  {Zoglauer}, {An}, {Bachetti}, {Barri{\`e}re}, {Bellm}, {Bhalerao},
  {Brejnholt}, {Fuerst}, {Liebe}, {Markwardt}, {Nynka}, {Vogel}, {Walton},
  {Wik}, {Alexander}, {Cominsky}, {Hornschemeier}, {Hornstrup}, {Kaspi},
  {Madejski}, {Matt}, {Molendi}, {Smith}, {Tomsick}, {Ajello}, {Ballantyne},
  {Balokovi{\'c}}, {Barret}, {Bauer}, {Blandford}, {Brandt}, {Brenneman},
  {Chiang}, {Chakrabarty}, {Chenevez}, {Comastri}, {Dufour}, {Elvis}, {Fabian},
  {Farrah}, {Fryer}, {Gotthelf}, {Grindlay}, {Helfand}, {Krivonos}, {Meier},
  {Miller}, {Natalucci}, {Ogle}, {Ofek}, {Ptak}, {Reynolds}, {Rigby},
  {Tagliaferri}, {Thorsett}, {Treister}, \& {Urry}}]{Harrison13}
{Harrison}, F.~A., {Craig}, W.~W., {Christensen}, F.~E., {et~al.} 2013, \apj,
  770, 103, \dodoi{10.1088/0004-637X/770/2/103}

\bibitem[{{HI4PI Collaboration} {et~al.}(2016){HI4PI Collaboration}, {Ben
  Bekhti}, {Fl{\"o}er}, {Keller}, {Kerp}, {Lenz}, {Winkel}, {Bailin},
  {Calabretta}, {Dedes}, {Ford}, {Gibson}, {Haud}, {Janowiecki}, {Kalberla},
  {Lockman}, {McClure-Griffiths}, {Murphy}, {Nakanishi}, {Pisano}, \&
  {Staveley-Smith}}]{HI4PI}
{HI4PI Collaboration}, {Ben Bekhti}, N., {Fl{\"o}er}, L., {et~al.} 2016, \aap,
  594, A116, \dodoi{10.1051/0004-6361/201629178}

\bibitem[{Hunter(2007)}]{Hunter07}
Hunter, J.~D. 2007, Computing In Science \& Engineering, 9, 90,
  \dodoi{10.1109/MCSE.2007.55}

\bibitem[{{Iwasawa} {et~al.}(2000){Iwasawa}, {Fabian}, {Almaini}, {Lira},
  {Lawrence}, {Hayashida}, \& {Inoue}}]{Iwasawa00}
{Iwasawa}, K., {Fabian}, A.~C., {Almaini}, O., {et~al.} 2000, \mnras, 318, 879,
  \dodoi{10.1046/j.1365-8711.2000.03810.x}

\bibitem[{{Iwasawa} {et~al.}(2010){Iwasawa}, {Tanaka}, \& {Gallo}}]{Iwasawa10}
{Iwasawa}, K., {Tanaka}, Y., \& {Gallo}, L.~C. 2010, \aap, 514, A58,
  \dodoi{10.1051/0004-6361/200912431}

\bibitem[{{Jansen} {et~al.}(2001){Jansen}, {Lumb}, {Altieri}, {Clavel}, {Ehle},
  {Erd}, {Gabriel}, {Guainazzi}, {Gondoin}, {Much}, {Munoz}, {Santos},
  {Schartel}, {Texier}, \& {Vacanti}}]{Jans01}
{Jansen}, F., {Lumb}, D., {Altieri}, B., {et~al.} 2001, \aap, 365, L1,
  \dodoi{10.1051/0004-6361:20000036}

\bibitem[{{Kaastra} \& {Bleeker}(2016)}]{Kaastra16}
{Kaastra}, J.~S., \& {Bleeker}, J.~A.~M. 2016, \aap, 587, A151,
  \dodoi{10.1051/0004-6361/201527395}

\bibitem[{{Kaastra} {et~al.}(2014){Kaastra}, {Kriss}, {Cappi}, {Mehdipour},
  {Petrucci}, {Steenbrugge}, {Arav}, {Behar}, {Bianchi}, {Boissay}, {Brand
  uardi-Raymont}, {Chamberlain}, {Costantini}, {Ely}, {Ebrero}, {Di Gesu},
  {Harrison}, {Kaspi}, {Malzac}, {De Marco}, {Matt}, {Nand ra}, {Paltani},
  {Person}, {Peterson}, {Pinto}, {Ponti}, {Nu{\~n}ez}, {De Rosa}, {Seta},
  {Ursini}, {de Vries}, {Walton}, \& {Whewell}}]{Kaastra14}
{Kaastra}, J.~S., {Kriss}, G.~A., {Cappi}, M., {et~al.} 2014, Science, 345, 64,
  \dodoi{10.1126/science.1253787}

\bibitem[{{Kammoun} {et~al.}(2018){Kammoun}, {Marin}, {Dov{\v{c}}iak},
  {Nardini}, {Risaliti}, \& {Sanfrutos}}]{Kammoun18}
{Kammoun}, E.~S., {Marin}, F., {Dov{\v{c}}iak}, M., {et~al.} 2018, \mnras, 480,
  3243, \dodoi{10.1093/mnras/sty2084}

\bibitem[{{Kammoun} {et~al.}(2019){Kammoun}, {Papadakis}, \&
  {Do{\v{c}}ciak}}]{Kammoun19}
{Kammoun}, E.~S., {Papadakis}, I.~E., \& {Do{\v{c}}ciak}, M. 2019, \apjl,
  arXiv:1906.07692.
\newblock \doarXiv{1906.07692}

\bibitem[{{Kara} {et~al.}(2016){Kara}, {Alston}, {Fabian}, {Cackett}, {Uttley},
  {Reynolds}, \& {Zoghbi}}]{Kara16}
{Kara}, E., {Alston}, W.~N., {Fabian}, A.~C., {et~al.} 2016, \mnras, 462, 511,
  \dodoi{10.1093/mnras/stw1695}

\bibitem[{{Karachentsev} \& {Drozdovsky}(1998)}]{Karachentsev98}
{Karachentsev}, I.~D., \& {Drozdovsky}, I.~O. 1998, \aaps, 131, 1,
  \dodoi{10.1051/aas:1998246}

\bibitem[{{Koliopanos} {et~al.}(2017){Koliopanos}, {Ciambur}, {Graham}, {Webb},
  {Coriat}, {Mutlu-Pakdil}, {Davis}, {Godet}, {Barret}, \&
  {Seigar}}]{Koliopanos17}
{Koliopanos}, F., {Ciambur}, B.~C., {Graham}, A.~W., {et~al.} 2017, \aap, 601,
  A20, \dodoi{10.1051/0004-6361/201630061}

\bibitem[{{Lira} {et~al.}(1999){Lira}, {Lawrence}, {O'Brien}, {Johnson},
  {Terlevich}, \& {Bannister}}]{Lira99}
{Lira}, P., {Lawrence}, A., {O'Brien}, P., {et~al.} 1999, \mnras, 305, 109,
  \dodoi{10.1046/j.1365-8711.1999.02388.x}

\bibitem[{{Matzeu} {et~al.}(2016){Matzeu}, {Reeves}, {Nardini}, {Braito},
  {Costa}, {Tombesi}, \& {Gofford}}]{Matzeu16}
{Matzeu}, G.~A., {Reeves}, J.~N., {Nardini}, E., {et~al.} 2016, \mnras, 458,
  1311, \dodoi{10.1093/mnras/stw354}

\bibitem[{{McHardy} {et~al.}(2006){McHardy}, {Koerding}, {Knigge}, {Uttley}, \&
  {Fender}}]{McHardy06}
{McHardy}, I.~M., {Koerding}, E., {Knigge}, C., {Uttley}, P., \& {Fender},
  R.~P. 2006, \nat, 444, 730, \dodoi{10.1038/nature05389}

\bibitem[{{McHardy} {et~al.}(2016){McHardy}, {Connolly}, {Peterson}, {Bieryla},
  {Chand}, {Elvis}, {Emmanoulopoulos}, {Falco}, {Gandhi}, \&
  {Kaspi}}]{McHardy16}
{McHardy}, I.~M., {Connolly}, S.~D., {Peterson}, B.~M., {et~al.} 2016,
  Astronomische Nachrichten, 337, 500, \dodoi{10.1002/asna.201612337}

\bibitem[{{Mezcua} {et~al.}(2018){Mezcua}, {Civano}, {Marchesi}, {Suh},
  {Fabbiano}, \& {Volonteri}}]{Mezcua18}
{Mezcua}, M., {Civano}, F., {Marchesi}, S., {et~al.} 2018, \mnras, 478, 2576,
  \dodoi{10.1093/mnras/sty1163}

\bibitem[{{Miniutti} {et~al.}(2014){Miniutti}, {Sanfrutos}, {Beuchert},
  {Ag{\'\i}s-Gonz{\'a}lez}, {Longinotti}, {Piconcelli}, {Krongold},
  {Guainazzi}, {Bianchi}, {Matt}, \& {Jim{\'e}nez-Bail{\'o}n}}]{Miniutti14}
{Miniutti}, G., {Sanfrutos}, M., {Beuchert}, T., {et~al.} 2014, \mnras, 437,
  1776, \dodoi{10.1093/mnras/stt2005}

\bibitem[{{Moran} {et~al.}(2005){Moran}, {Eracleous}, {Leighly}, {Chartas},
  {Filippenko}, {Ho}, \& {Blanco}}]{Moran05}
{Moran}, E.~C., {Eracleous}, M., {Leighly}, K.~M., {et~al.} 2005, \aj, 129,
  2108, \dodoi{10.1086/429522}

\bibitem[{{Nandra} {et~al.}(2007){Nandra}, {O'Neill}, {George}, \&
  {Reeves}}]{Nandra07}
{Nandra}, K., {O'Neill}, P.~M., {George}, I.~M., \& {Reeves}, J.~N. 2007,
  \mnras, 382, 194, \dodoi{10.1111/j.1365-2966.2007.12331.x}

\bibitem[{{Nardini} \& {Risaliti}(2011)}]{Nardini11}
{Nardini}, E., \& {Risaliti}, G. 2011, \mnras, 417, 2571,
  \dodoi{10.1111/j.1365-2966.2011.19423.x}

\bibitem[{{Nasa High Energy Astrophysics Science Archive Research Center
  (Heasarc)}(2014)}]{HEASoft}
{Nasa High Energy Astrophysics Science Archive Research Center (Heasarc)}.
  2014, {HEAsoft: Unified Release of FTOOLS and XANADU}, Astrophysics Source
  Code Library.
\newblock \doeprint{1408.004}

\bibitem[{{Parker} {et~al.}(2015){Parker}, {Fabian}, {Matt}, {Koljonen},
  {Kara}, {Alston}, {Walton}, {Marinucci}, {Brenneman}, \&
  {Risaliti}}]{Parker15}
{Parker}, M.~L., {Fabian}, A.~C., {Matt}, G., {et~al.} 2015, \mnras, 447, 72,
  \dodoi{10.1093/mnras/stu2424}

\bibitem[{{Peterson} {et~al.}(2005){Peterson}, {Bentz}, {Desroches},
  {Filippenko}, {Ho}, {Kaspi}, {Laor}, {Maoz}, {Moran}, \&
  {Pogge}}]{Peterson05}
{Peterson}, B.~M., {Bentz}, M.~C., {Desroches}, L.-B., {et~al.} 2005, \apj,
  632, 799, \dodoi{10.1086/444494}

\bibitem[{{Reeves} {et~al.}(2008){Reeves}, {Done}, {Pounds}, {Terashima},
  {Hayashida}, {Anabuki}, {Uchino}, \& {Turner}}]{Reeves08}
{Reeves}, J., {Done}, C., {Pounds}, K., {et~al.} 2008, \mnras, 385, L108,
  \dodoi{10.1111/j.1745-3933.2008.00443.x}

\bibitem[{{Risaliti} {et~al.}(2005){Risaliti}, {Elvis}, {Fabbiano}, {Baldi}, \&
  {Zezas}}]{Risaliti05}
{Risaliti}, G., {Elvis}, M., {Fabbiano}, G., {Baldi}, A., \& {Zezas}, A. 2005,
  \apjl, 623, L93, \dodoi{10.1086/430252}

\bibitem[{{Schurch} \& {Warwick}(2002)}]{Schurch02}
{Schurch}, N.~J., \& {Warwick}, R.~S. 2002, \mnras, 334, 811,
  \dodoi{10.1046/j.1365-8711.2002.05546.x}

\bibitem[{{Silva} {et~al.}(2016){Silva}, {Uttley}, \& {Costantini}}]{Silva16}
{Silva}, C.~V., {Uttley}, P., \& {Costantini}, E. 2016, \aap, 596, A79,
  \dodoi{10.1051/0004-6361/201628555}

\bibitem[{{Smith} {et~al.}(2001){Smith}, {Brickhouse}, {Liedahl}, \&
  {Raymond}}]{Smith01}
{Smith}, R.~K., {Brickhouse}, N.~S., {Liedahl}, D.~A., \& {Raymond}, J.~C.
  2001, \apjl, 556, L91, \dodoi{10.1086/322992}

\bibitem[{{Str{\"u}der} {et~al.}(2001){Str{\"u}der}, {Briel}, {Dennerl},
  {Hartmann}, {Kendziorra}, {Meidinger}, {Pfeffermann}, {Reppin}, {Aschenbach},
  {Bornemann}, {Br{\"a}uninger}, {Burkert}, {Elender}, {Freyberg}, {Haberl},
  {Hartner}, {Heuschmann}, {Hippmann}, {Kastelic}, {Kemmer}, {Kettenring},
  {Kink}, {Krause}, {M{\"u}ller}, {Oppitz}, {Pietsch}, {Popp}, {Predehl},
  {Read}, {Stephan}, {St{\"o}tter}, {Tr{\"u}mper}, {Holl}, {Kemmer}, {Soltau},
  {St{\"o}tter}, {Weber}, {Weichert}, {von Zanthier}, {Carathanassis}, {Lutz},
  {Richter}, {Solc}, {B{\"o}ttcher}, {Kuster}, {Staubert}, {Abbey}, {Holland},
  {Turner}, {Balasini}, {Bignami}, {La Palombara}, {Villa}, {Buttler},
  {Gianini}, {Lain{\'e}}, {Lumb}, \& {Dhez}}]{Stru01}
{Str{\"u}der}, L., {Briel}, U., {Dennerl}, K., {et~al.} 2001, \aap, 365, L18,
  \dodoi{10.1051/0004-6361:20000066}

\bibitem[{{Tashiro} {et~al.}(2018){Tashiro}, {Maejima}, {Toda}, {Kelley},
  {Reichenthal}, {Lobell}, {Petre}, {Guainazzi}, {Costantini}, \&
  {Edison}}]{Xrism}
{Tashiro}, M., {Maejima}, H., {Toda}, K., {et~al.} 2018, in Society of
  Photo-Optical Instrumentation Engineers (SPIE) Conference Series, Vol. 10699,
  Space Telescopes and Instrumentation 2018: Ultraviolet to Gamma Ray, 1069922

\bibitem[{{Vaughan}(2010)}]{Vaughan10}
{Vaughan}, S. 2010, \mnras, 402, 307, \dodoi{10.1111/j.1365-2966.2009.15868.x}

\bibitem[{{Vaughan} {et~al.}(2003){Vaughan}, {Edelson}, {Warwick}, \&
  {Uttley}}]{Vaughan03}
{Vaughan}, S., {Edelson}, R., {Warwick}, R.~S., \& {Uttley}, P. 2003, \mnras,
  345, 1271, \dodoi{10.1046/j.1365-2966.2003.07042.x}

\bibitem[{{Vaughan} {et~al.}(2005){Vaughan}, {Iwasawa}, {Fabian}, \&
  {Hayashida}}]{Vaughan05}
{Vaughan}, S., {Iwasawa}, K., {Fabian}, A.~C., \& {Hayashida}, K. 2005, \mnras,
  356, 524, \dodoi{10.1111/j.1365-2966.2004.08463.x}

\bibitem[{{Whittle}(1953)}]{Whittle53}
{Whittle}, P. 1953, Arkiv for Matematik, 2, 423, \dodoi{10.1007/BF02590998}

\bibitem[{{Woo} {et~al.}(2019){Woo}, {Cho}, {Gallo}, {Hodges-Kluck}, {Le},
  {Shin}, {Son}, \& {Horst}}]{Woo19}
{Woo}, J.-H., {Cho}, H., {Gallo}, E., {et~al.} 2019, arXiv e-prints,
  arXiv:1905.00145.
\newblock \doarXiv{1905.00145}

\bibitem[{{Zhang} {et~al.}(2019){Zhang}, {Santangelo}, {Feroci}, {Xu}, {Lu},
  {Chen}, {Feng}, {Zhang}, {Brandt}, \& {Hernanz}}]{eXTP19}
{Zhang}, S., {Santangelo}, A., {Feroci}, M., {et~al.} 2019, Science China
  Physics, Mechanics, and Astronomy, 62, 29502,
  \dodoi{10.1007/s11433-018-9309-2}

\bibitem[{{Zoghbi} {et~al.}(2019){Zoghbi}, {Miller}, \& {Cackett}}]{Zoghbi19}
{Zoghbi}, A., {Miller}, J., \& {Cackett}, E. 2019, ApJ, arXiv:1908.09862.
\newblock \doarXiv{1908.09862}

\end{thebibliography}
\end{document}